\documentclass[%
 reprint,
nofootinbib,
 amsmath,amssymb,
 aps,
]{revtex4-2}

\usepackage{graphicx}
\usepackage{dcolumn}
\usepackage{bm}
\usepackage{xcolor}
\usepackage[normalem]{ulem} 
\usepackage[hidelinks]{hyperref}

\newcommand{\mbf}[1]{\mathbf{#1}}

\DeclareMathAlphabet\mathbfcal{OMS}{cmsy}{b}{n}
\DeclareMathOperator{\dprime}{\prime \prime}

\begin{document}

\preprint{APS/123-QED}

\title{Demonstration of hybrid foreground removal on CHIME data}

\author{Haochen~Wang$^{1,2}$}
\email[]{hcwang96@mit.edu}
\author{Kiyoshi~Masui$^{1,2}$}
\author{Kevin~Bandura$^{3,4}$}
\author{Arnab~Chakraborty$^{5,6}$}
\author{Matt Dobbs$^{5,6}$}
\author{Simon~Foreman$^{7}$}
\author{Liam~Gray$^{8}$}
\author{Mark~Halpern$^{8}$}
\author{Albin~Joseph$^{7}$}
\author{Joshua~MacEachern$^{8}$}
\author{Juan~Mena-Parra$^{9, 10}$}
\author{Kyle~Miller$^{5,6}$}
\author{Laura~Newburgh$^{11}$}
\author{Sourabh~Paul$^{12, 5}$}
\author{Alex~Reda$^{11}$}
\author{Pranav~Sanghavi$^{11}$}
\author{Seth~Siegel$^{13, 5, 6}$}
\author{Dallas~Wulf$^{5, 6}$}
\affiliation{$^1$ Department of Physics, Massachusetts Institute of Technology, 77 Massachusetts Avenue Cambridge, MA 02139, USA}
\affiliation{$^2$ MIT Kavli Institute for Astrophysics and Space Research, Massachusetts Institute of Technology, 77 Massachusetts Avenue Cambridge, MA 02139, USA}
\affiliation{$^3$ Lane Department of Computer Science and Electrical Engineering, 1220 Evansdale Drive, PO Box 6109, Morgantown, WV 26506, USA}
\affiliation{$^4$ Center for Gravitational Waves and Cosmology, West Virginia University, Chestnut Ridge Research Building, Morgantown, WV 26505, USA}
\affiliation{$^5$ Department of Physics, McGill University, 3600 rue University, Montr\'eal, QC H3A 2T8, Canada}
\affiliation{$^6$ Trottier Space Institute, McGill University, 3550 rue University, Montr\'eal, QC H3A 2A7, Canada}
\affiliation{$^7$ Department of Physics, Arizona State University, Tempe, AZ, USA}
\affiliation{$^8$ Department of Physics and Astronomy, University of British Columbia, 6224 Agricultural Road, Vancouver, BC V6T 1Z1 Canada}
\affiliation{$^9$ Dunlap Institute for Astronomy \& Astrophysics, University of Toronto, 50 St.~George Street, Toronto, ON M5S 3H4, Canada}
\affiliation{$^{10}$ David A.~Dunlap Department of Astronomy \& Astrophysics, University of Toronto, 50 St.~George Street, Toronto, ON M5S 3H4, Canada}
\affiliation{$^{11}$ Department of Physics, Yale University, New Haven, CT 06520, USA}
\affiliation{$^{12}$ Jodrell Bank Centre for Astrophysics, School of Physics and Astronomy, The University of Manchester, Manchester M13 9PL, UK}
\affiliation{$^{13}$ Perimeter Institute for Theoretical Physics, 31 Caroline Street N, Waterloo, ON N25 2YL, Canada}

\date{\today}

\begin{abstract}

The main challenge of 21 cm cosmology experiments is astrophysical foregrounds which are difficult to separate from the signal due to telescope systematics. An earlier study has shown that foreground residuals induced by antenna gain errors can be estimated and subtracted using the hybrid foreground residual subtraction (HyFoReS) technique which relies on cross-correlating linearly filtered data. In this paper, we apply a similar technique to the CHIME stacking analysis to subtract beam-induced foreground contamination. Using a linear high-pass delay filter for foreground suppression, the CHIME collaboration reported a $11.1\sigma$ detection in the 21 cm signal stacked on eBOSS quasar locations, despite foreground residual contamination mostly due to the instrument chromatic transfer function. We cross-correlate the foreground-dominated data at low delay with the contaminated signal at high delay to estimate residual foregrounds and subtract them from the signal. We find foreground residual subtraction can improve the signal-to-noise ratio of the stacked 21 cm signal by $ 10 - 20\%$ after the delay foreground filter, although some of the improvement can also be achieved with an alternative flagging technique. We have shown that it is possible to use HyFoReS to reduce beam-induced foreground contamination, benefiting the analysis of the HI auto power spectrum with CHIME and enabling the recovery of large scale modes.

\end{abstract}

\maketitle

\section{introduction}

The hyperfine transition of neutral hydrogen atoms (HI) emits photons at 21 cm wavelength, which can be observed to map out the large-scale structures (LSS) of the universe. One promising technique to do so is hydrogen intensity mapping \citep{chang_pen, Loeb}. In this approach, radio telescopes observe accumulated 21 cm emissions from many galaxies to perform a coarse-grained survey. Compared with traditional galaxy surveys, intensity mapping has the advantages of rapid mapping speed and wide redshift coverage, allowing observation of large volumes of the universe through most of cosmic history. 

Many 21 cm intensity mapping experiments are underway. Low and intermediate-redshift surveys, such as CHIME \citep{chime_overview}, MeerKAT \citep{meerkat}, and uGMRT \citep{ugmrt}, aim to measure hydrogen mass density and baryon acoustic oscillations (BAO) from the underlying matter distributions at $z < 4$, while high-redshift observations, such as HERA \citep{HERA_2017a}, LOFAR \citep{LOFAR_2017}, and MWA \citep{MWA_2019}, probe ionization fractions and constrain reionization models around $z \sim 6 - 10$. These experiments have produced many notable results, including cross-correlations with other LSS surveys \citep{chime_lyman, chime_stacking, GBT_cross, meerkat_cross, Li_2021, Anderson_2018, kiyo_survey, Chang_2010}, upper-limits on auto-correlation of 21 cm signal \citep{Switzer_2013, HERA_limit, ugmrt}, and a first detection of auto-correlation \citep{paul2023detection}. 

However, current 21 cm intensity mapping surveys still face the challenge of bright astrophysical foregrounds consisting mostly of Galactic synchrotron emissions and extra-galactic point sources, which are 3 to 5 orders of magnitude brighter than the 21 cm signal \citep{first_mmode}. Linear foreground filtering methods, such as the delay filter \citep{DAYENU} and Karhunen-Loève (KL) eigenmode projection \citep{KL_cite, mmode}, can in principle separate foregrounds from the signal by utilizing the smooth spectral characters of foregrounds. However, this class of techniques is highly sensitive to unknown instrumental responses and calibration errors, refered to as ``telescope systematics", and can leave bright foreground residuals in the filtered signal. Nonlinear methods, such as those based on principal component analysis (PCA) \citep{GBT_cross, Anderson_2018, kiyo_survey,Chang_2010}, can remove foregrounds by zeroing out the most dominant modes from the data but risk significant signal loss that is difficult to characterize \citep{kiyo_survey}.

To mitigate foreground contamination, most 21 cm experiments targeting lower redshifts ($z < 4$) have relied on cross-correlation with other types of LSS surveys. In particular, CHIME cross-correlated its 21 cm measurement with luminous red galaxies (LRG), emission line galaxies (ELG), and quasars (QSO) from the eBOSS clustering catalogs \citep{chime_stacking}. By stacking the 21 cm intensity map on the spatial and spectral locations of eBOSS catalog objects, the CHIME stacking analysis detected 21 cm signal with high significance and placed constraints on the cosmic abundance of HI. A similar cross-correlation was also performed with the eBOSS Lyman-alpha measurement \citep{chime_lyman}. However, due to systematics such as uncharacterized features from CHIME's primary beam, foregrounds leak beyond the subspace which they intrinsically occupy in the data and contaminate the 21 cm measurement. As a result, an aggressive high-pass delay foreground filter had to be applied in these studies, which removed all sensitivity to BAO from the data and limited the measurement to small and nonlinear physical scales. 

New foreground mitigation techniques have recently emerged to address the issues of foreground residual contamination and signal loss that undermine the traditional linear and nonlinear foreground removal methods. In particular, a previous study has proposed a family of new algorithms called HyFoReS (Hybrid Foreground Residual Subtraction), which first uses a traditional linear foreground filter to produce estimated foregrounds and signal, and then cross-correlates the foreground and signal estimates to draw out foreground residuals and subtract them from the estimated signal \citep{haochen_first_paper}. The technique was shown to reduce foreground leakage from antenna gain perturbations and improve the 21 cm power spectrum measurement in simulations of a small antenna array. 

In this work, we apply HyFoReS to CHIME data to remove beam-induced foreground residuals. We show that HyFoReS can improve the signal-to-noise ratio of the stacked 21 cm signal by $10\%$ to $20\%$ after the delay foreground filter has been applied. With this improvement, the method can afford using a less aggressive delay foreground filter to preserve more large-scale information in the data and potentially benefit the measurement of HI auto power spectrum in the future. 

We organize the paper as follows. We first summarize the CHIME stacking analysis in Section~\ref{sec:motiv}, highlighting some of the challenges in foreground filtering which motivated this study. In Section~\ref{sec:form}, we give a brief review on the general formalism of HyFoReS and then adapt it to solving the foreground leakage problem in the CHIME stacking data. We describe how we implement HyFoReS in the CHIME stacking pipeline in Section~\ref{sec:app} and explain how we quantify the detection significance. In Section~\ref{sec:res}, we present the result by fitting a 21 cm signal template to the stacked data and compare the signal-to-noise ratios before and after applying HyFoReS. We further explore combining HyFoReS with an outlier mask employed in the original stacking analysis and experiment with using lower cutoffs for the delay filter. In Section~\ref{sec:dis}, we address the assumptions and subtleties of HyFoReS and comment on its potential as well as improvements expected in the future. Finally, we present our conclusions in Section~\ref{sec:conclu}.

\section{motivation} \label{sec:motiv}

The Canadian Hydrogen Intensity Mapping Experiment (CHIME) is a transit radio telescope located at the Dominion Radio Astrophysical Observatory in British Columbia Canada. The telescope is an interferometer consisting of four parallel 100m $\times$ 20m cylindrical reflectors with a total of 1024 dual-polarization feeds, covering the 400 MHz to 800 MHz band. The CHIME stacking analysis \citep{chime_stacking} used observation data from 102 nights to produce an HI intensity map, which was foreground filtered and then stacked on the angular and spectral locations of luminous red galaxies (LRG), emission line galaxies (ELG), and quasars (QSO) from the eBOSS clustering catalogs. The analysis achieved decisive detection of neutral hydrogen using all three tracers, with significance of 7.1$\sigma$, 5.7$\sigma$, and 11.1$\sigma$, respectively, by likelihood-ratio tests and placed constraints on the effective clustering amplitude of neutral hydrogen.

The CHIME HI intensity map was made with inter-cylinder baselines (i.e.\ pairs of feeds occupying different cylinders), which have low sensitivity to diffuse foreground emissions from the Galaxy and noise cross-talk between feeds. In addition, a linear high-pass delay filter was constructed using the DAYENU technique \citep{DAYENU} to further mitigate foregrounds. Foregrounds have intrinsically smooth frequency spectra. These smooth frequency modes translate to low-delay modes centered narrowly around 0 in delay space (Fourier conjugate of the frequency domain). The DAYENU delay filter removes foregrounds by discarding data below a certain delay cutoff $\tau_{\text{cutoff}}$ while mitigating subtle effects from band-limited and irregular sampling of the data.

CHIME's systematics, such as the uncertain primary beam, introduce more complex frequency dependence to the otherwise smooth foregrounds, causing foreground power to leak into high delay signal modes. To suppress foreground leakage, the delay cutoff was chosen to be between 130 and 190 nanoseconds, depending on declination and polarization (shown in Fig.~\ref{fig:cutoffs}). This was referred to as the ``relaxed cutoff" in the stacking analysis. This choice is much higher than where the intrinsic foregrounds occupy, causing signal loss at large scales along the line of sight (namely, the low $k_\parallel$ modes). As a result, the post-filtered data is only sensitive to structures with comoving wavenumber $k_\parallel$ $>$ 0.3 $h$ $\text{Mpc}^{-1}$, which excludes any sensitivity to BAO from the data. In addition, the accessible physical scales are mostly too small for perturbative analysis, making the result difficult to interpret and vulnerable to uncertainties in the modelling of nonlinear clustering.


To improve the CHIME stacking analysis, we must address the effect of systematics on the data. The most significant source of systematics for CHIME is uncertainties in the primary beam response \citep{chime_overview, chime_stacking}. Although the stacking analysis estimated the primary beam with point source measurements \citep{chime_stacking}, the deconvolution procedure in the analysis was unable to deconvolve all of the beam response from the data (see a detailed discussion in Appendix~\ref{app:decon}). In addition, there are uncertainties in the beam model. Specifically, uncharacterized beam chromaticity (the beam's dependence on frequency) would imprint the beam's frequency structure on the foregrounds even after deconvolution, breaking the assumption of the high-pass delay foreground filter and causing foreground residuals in the filtered data.


\begin{figure*}
\includegraphics[scale=0.235]{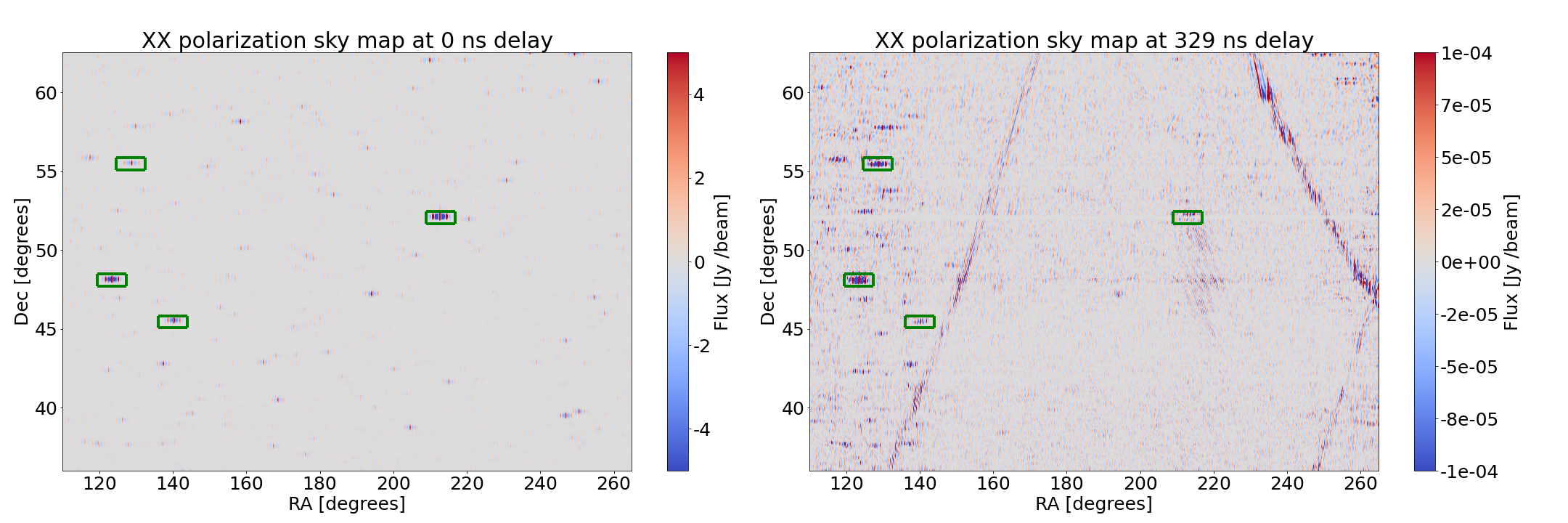}
\caption{\label{fig:low_vs_high} Comparison of the XX polarization sky maps at 0 and 329 nano-second delays. Left panel: sky map at 0 delay after deconvolution but before foreground filtering, which is also the estimated foregrounds used as inputs for HyFoReS. Several point sources are visible with four of them highlighted within green boxes for visualization. Right panel: sky map at 329 ns delay after deconvolution and foreground filtering. Bright point sources cause visible leakage, with the same four regions highlighted in the green boxes as the ones in the left panel. The bright U-shaped stripes on the left side, upper right corner, and lower right corner of the right panel come from the far sidelobes due to Taurus A, Cygnus A, and Virgo A, respectively, which will be masked in subsequent analysis.}
\end{figure*}

These foreground residuals can be seen in the filtered signal, as demonstrated by the right panel of Fig.~\ref{fig:low_vs_high}, which shows the CHIME stacking data at a high delay (329 ns) above the relaxed delay cutoff after the data has already been deconvolved (Eq.~[\ref{eq:deconv}]) and foreground filtered. Comparing with the zero delay foreground map on the left, we observe that some point sources from the foreground map leak into the signal map, with some spreading along both RA and Dec directions due to the instrument response. The procedure for obtaining the map as a function of delay, which we refer to as the delay spectrum, is described in Appendix~\ref{app:dspec}.



Foreground residuals wash out the 21 cm signal and contribute to excess noise after stacking. This forces the delay foreground filter to be more aggressive at the cost of losing signal and sensitivity to large-scale structures. The purpose of this study is twofold. First, we want to use HyFoReS to remove beam-induced foreground residuals from the CHIME stacking data and improve the detection significance. Second, we explore the possibility of using a less aggressive delay foreground filter to preserve more signal and large-scale structure information. This particular application of HyFoReS also serves as a proof of concept of using this algorithm to mitigate beam errors for 21 cm intensity mapping experiments in general. 


\section{formalism} \label{sec:form}

In this study, we use HyFoReS to remove beam-induced foreground leakage in the CHIME stacking data. We first review the general formalism of the algorithm in Subsection~\ref{subsec:form_rev}. We then adapt the formalism to the CHIME stacking analysis in Subsection~\ref{subsec:adapt}.

\subsection{Review of HyFoReS} \label{subsec:form_rev}

The systematic-robust hybrid foreground residual subtraction (HyFoReS) algorithm was first derived in \citep{haochen_first_paper}. The essence of the algorithm is to first use a traditional linear foreground filter to obtain a rough signal and foreground estimate. Because of telescope systematics, the signal estimate is dominated by foreground residuals. The algorithm then cross-correlates the signal estimate with the foreground estimate to draw out the foreground residual contamination and then subtracts it from the signal estimate.

We use $\bm{d}$, $\bm{s}$, and $\bm{f}$ to represent the data, signal, and foregrounds, respectively, in the form of vectors. Assuming telescope systematics are parametrizable and act like multiplicative perturbation factors on the signal and foregrounds, we can model our data as 
\begin{equation} \label{eq:data_form}
    \bm{d} = (\mbf{I} + \mbf{G})(\bm{s} + \bm{f}),
\end{equation}
where
\begin{equation} \label{eq:G_form}
    \mbf{G} = \sum_i g_i \mbf{\Gamma_i}
\end{equation}
is a perturbation matrix that multiplies the signal and foregrounds with a set of systematics $g_i$, and $\mbf{I}$ is the identiy matrix. The matrix $\mbf{\Gamma_i}$ serves as a derivative matrix for each perturbation $g_i$ and determines which subspace of the data each perturbation $g_i$ affects. Note that Eq.~(\ref{eq:data_form}) does not include noise. We expect the noise term $\bm{n}$ to propagate through the foreground subtraction procedure like the signal $\bm{s}$ and perturbed signal $\mbf{G}\bm{s}$, assuming the foreground filter removes foregrounds but preserves both signal and noise and that the foreground terms $\bm{f}$ and $\mbf{G}\bm{f}$ in Eq.~(\ref{eq:data_form}) dominate over the signal and noise terms. The noise bias can be later removed in the power spectrum estimation step, for example, by cross-correlating data from two seasons of observation over the same sky.

We can use a traditional linear foreground filter $\mbf{K}$, (such as the delay filter, KL filter, or any other linear filter), to obtain an estimated signal
\begin{equation} \label{eq:s_hat_form}
    \bm{\hat{s}} = \mbf{K}\bm{d} \approx \mbf{KG}\bm{f}.    
\end{equation}
Note that the estimated signal is dominated by foreground residuals ($\mbf{KG}\bm{f}$), assuming that the foreground filter sufficiently removes the intrinsic foregrounds (i.e., $\mbf{K}\bm{f} \ll \bm{s}$). Similarly, we can obtain the estimated foregrounds
\begin{equation} \label{eq:f_hat_form}
    \bm{\hat{f}} = \mbf{A}\bm{d} \approx \bm{f},
\end{equation}
where $\mbf{A}$ can be any linear filter that gives a foreground estimate. (For example, we can choose $\mbf{A} = \mbf{I} - \mbf{K}$, which means using the signal subtracted data as the foreground estimate, or simply use $\mbf{A} = \mbf{I}$ when the data is already foreground dominated).

By the formalism derived in \citep{haochen_first_paper}, the telescope systematics $g_i$ can be estimated by cross-correlating the estimated signal and foregrounds
\begin{equation} \label{eq:y_hat_form}
    \bm{\hat{y}}_i = \frac{\bm{\hat{f}}^\dagger\mbf{E_i}^\dagger\bm{\hat{s}}}{\bm{\hat{f}}^\dagger\mbf{D_i}\bm{\hat{f}}},
\end{equation}
where $\mbf{E_i}$ is a matrix that determines over which subspace the estimated signal and foreground should be cross-correlated for estimating each systematics $g_i$, and $\mbf{D_i}$ entails how the normalization is done. Previous work \citep{haochen_first_paper} has shown that having $\mbf{E_i} \propto \mbf{\Gamma_i}$ and $\mbf{D_i} = \mbf{E_i}^\dagger \mbf{E_i}$ are good choices in general. The estimated systematics $\bm{\hat{y}}_i$ in Eq.~(\ref{eq:y_hat_form}) is a linear combination of the true systematics $\bm{g}_i$. This can be easily seen by plugging Eq.~(\ref{eq:G_form}) into (\ref{eq:s_hat_form}) and then (\ref{eq:s_hat_form}) into~(\ref{eq:y_hat_form}) , giving
\begin{equation} \label{eq:window_form}
    \bm{\hat{y}}_i = \sum_{i^\prime} \frac{\bm{\hat{f}}^\dagger\mbf{E_i}^\dagger\mbf{K}\mbf{\Gamma_{i^\prime}}\bm{f}}{\bm{\hat{f}}^\dagger\mbf{D_i}\bm{\hat{f}}} g_{i^\prime} = \sum_{i^\prime} \mbf{W}_{i i^\prime}g_{i^\prime},
\end{equation}
where
\begin{equation} \label{eq:window_def_form}
    \mbf{W}_{i i^\prime} = \frac{\bm{\hat{f}}^\dagger\mbf{E_i}^\dagger\mbf{K}\mbf{\Gamma_{i^\prime}}\bm{f}}{\bm{\hat{f}}^\dagger\mbf{D_i}\bm{\hat{f}}}
\end{equation}
is the window matrix. Equation~(\ref{eq:window_def_form}) contains the true foregrounds $\bm{f}$ which we do not know. At the expense of introducing second order errors in the perturbations ($\propto \mbf{G}^2 \bm{f}$), we can compute Eq.~(\ref{eq:window_def_form}) using the estimated foreground $\bm{\hat{f}}$ in place of the true foregrounds $\bm{f}$, namely
\begin{equation} \label{eq:window_approx_form}
    \mbf{W}_{i i^\prime} \approx \frac{\bm{\hat{f}}^\dagger\mbf{E_i}^\dagger\mbf{K}\mbf{\Gamma_{i^\prime}}\bm{\hat{f}}}{\bm{\hat{f}}^\dagger\mbf{D_i}\bm{\hat{f}}}.
\end{equation}

Then we can obtain the unwindowed systematics estimate
\begin{equation}
    \bm{\hat{g}_i} = \sum_{i^\prime} \mbf{W}^+_{i i^\prime}\bm{\hat{y}}_{i^\prime},
\end{equation}
where $\mbf{W}^+_{i i^\prime}$ is the pseudo inverse of the window matrix from Eq.~(\ref{eq:window_approx_form}). With the systematics estimated, we can reconstruct the perturbation matrix from Eq.~(\ref{eq:G_form})
\begin{equation}
    \mbf{\hat{G}} = \sum_i \hat{g}_i \mbf{\Gamma_i}
\end{equation}
and subtract foreground residuals from the estimated signal to obtain the cleaned signal
\begin{equation}
    \bm{\tilde{s}} = \bm{\hat{s}} - \mbf{K \hat{G}}\bm{\hat{f}}. 
\end{equation}
Note that because of the approximation we used in Eq.~(\ref{eq:window_approx_form}), the cleaned signal is still contaminated by foregrounds coupled with second-order perturbations, namely
\begin{equation}
    \bm{\tilde{s}} \sim \bm{s} + \mbf{G}^2 \bm{f}.
\end{equation}
Telescopes are generally calibrated in real experiments. The systematics $\bm{\hat{g}_i}$, which arise from calibration errors and uncharacterized instrumental effects, are usually small (for example, at percent or sub-percent level). As long as the second order errors are not brighter than the signal, namely $\mbf{G}^2 \bm{f} < \bm{s}$, HyFoReS can sufficiently remove foreground residual contamination and recover the true signal.

\subsection{Adaptation to the CHIME stacking analysis} \label{subsec:adapt}

Now we adapt the general formalism of HyFoReS to the CHIME stacking analysis. As discussed in Section~\ref{sec:motiv}, most of foreground residuals in high delay maps originate from foregrounds in low delays. Since most foreground power is concentrated in the zero delay map, we assume that leakage in high delays comes purely from foregrounds in the zero delay map. We can then model the leakage using a parameter $b_{p d h}(\tau)$, which indicates how foreground sources at declination (Dec) $p$ in the zero delay map leak into a high delay map with delay $\tau$ at Dec $d$. Note that we have assumed stationarity in right ascension (RA) $\alpha$ as in \citep{mmode} and \citep{chime_stacking}, so the leakage parameter depends on $h \equiv \Delta \alpha$, the difference in RA between the zero and high delay maps. We can now model the high delay map as
\begin{equation} \label{eq:map_model}
    m_{d \alpha} (\tau) = s_{d \alpha}(\tau) + \sum_{p, h} b_{p d h}(\tau) m_{p(\alpha-h)}(0).
\end{equation}
The first term on the right hand side of Eq.~(\ref{eq:map_model}), $s_{d \alpha}(\tau)$, is the signal component of the map (with noise absorbed in it), and the second term is the foreground leakage. Note that the leakage parameters $b_{p d h}(\tau)$ serve as a Dec and delay-dependent convolution kernel that convolves foregrounds in the zero delay map and sends them to high delay maps.

To match the formalism from Section~\ref{subsec:form_rev}, we recognize the zero delay map as the estimated foregrounds and the high delay maps as the estimated signal, namely
\begin{align}
    & \hat{f}_{d \alpha} = m_{d \alpha} (0) \label{eq:f_hat_algo}\\
    & \hat{s}_{d \alpha} (\tau) = m_{d \alpha} (\tau). \label{eq:s_hat_algo}
\end{align}
To write Eq.~(\ref{eq:map_model}) in a compact matrix notation similar to Eq.~(\ref{eq:s_hat_form}), we define the derivative matrix
\begin{equation} \label{eq:gamma_def_apply}
    (\mbf{\Gamma}_{pdh})_{(d^\prime \alpha^\prime)(d^{\dprime} \alpha^{\dprime})} = \delta_{d d^\prime} \delta_{p d^{\dprime}} \delta_{\alpha^{\dprime} (\alpha^\prime - h)}
\end{equation}
for each parameter $b_{p d h}(\tau)$, and define the transfer matrix
\begin{equation}
    \mbf{B}(\tau) = \sum_{p,d,h} b_{pdh}(\tau) \mbf{\Gamma}_{pdh}.   
\end{equation}
Now Eq.~(\ref{eq:map_model}) can be written as
\begin{equation} \label{eq:s_hat_apply}
    \bm{\hat{s}} (\tau) = \bm{s}(\tau) + \mbf{B}(\tau) \bm{\hat{f}} \approx \mbf{B}(\tau) \bm{\hat{f}},
\end{equation}
where we can ignore the signal component for now given the assumption that foreground residuals dominate.

It is straightforward to apply the rest of the formalism from Section~\ref{subsec:form_rev}. Given the estimated signal $\bm{\hat{s}}$ and foregrounds $\bm{\hat{f}}$, we can estimate the leakage parameters in a way similar to Eq.~(\ref{eq:y_hat_form}),
\begin{equation} \label{eq:b_hat_def_pre}
    \hat{b}_{pdh} (\tau) = \frac{\bm{\hat{f}}^\dagger \mbf{E}_{pdh}^\dagger \bm{\hat{s}}(\tau)}{\bm{\hat{f}}^\dagger \mbf{D}_{pdh} \bm{\hat{f}}}.
\end{equation}
Choosing $\mbf{E}_{pdh} = \mbf{\Gamma}_{pdh}$ and $\mbf{D}_{pdh} = \mbf{\Gamma}_{pdh}^\dagger \mbf{\Gamma}_{pdh}$, Eq.~(\ref{eq:b_hat_def_pre}) becomes
\begin{equation} \label{eq:b_hat_def}
    \hat{b}_{pdh} (\tau) = \frac{\bm{\hat{f}}^\dagger \mbf{\Gamma}_{pdh}^\dagger \bm{\hat{s}}(\tau)}{\bm{\hat{f}}^\dagger \mbf{\Gamma}_{pdh}^\dagger \mbf{\Gamma}_{pdh} \bm{\hat{f}}}.
\end{equation}
Using the model of $\bm{\hat{s}}$ in Eq.~(\ref{eq:s_hat_apply}), we get
\begin{equation}
\begin{split}
    \hat{b}_{pdh} (\tau) & = \sum_{p^\prime, d^\prime, h^\prime} \frac{\bm{\hat{f}}^\dagger \mbf{\Gamma}_{pdh}^\dagger \mbf{\Gamma}_{p^\prime d^\prime h^\prime} \bm{\hat{f}}}{\bm{\hat{f}}^\dagger \mbf{\Gamma}_{pdh}^\dagger \mbf{\Gamma}_{pdh} \bm{\hat{f}}} b_{p^\prime d^\prime h^\prime} (\tau)\\
    & = \sum_{p^\prime, d^\prime, h^\prime} \mbf{W}_{(pdh)(p^\prime d^\prime h^\prime)} b_{p^\prime d^\prime h^\prime} (\tau),
\end{split}
\end{equation}
where
\begin{equation}
    \mbf{W}_{(pdh)(p^\prime d^\prime h^\prime)} = \frac{\bm{\hat{f}}^\dagger \mbf{\Gamma}_{pdh}^\dagger \mbf{\Gamma}_{p^\prime d^\prime h^\prime} \bm{\hat{f}}}{\bm{\hat{f}}^\dagger \mbf{\Gamma}_{pdh}^\dagger \mbf{\Gamma}_{pdh} \bm{\hat{f}}}
\end{equation}
is the window matrix. Using the definition of $\mbf{\Gamma}_{pdh}$ in Eq.~(\ref{eq:gamma_def_apply}), we find
\begin{equation} \label{eq:window_apply_explicit}
    \mbf{W}_{(pdh)(p^\prime d^\prime h^\prime)} = \frac{\sum_\alpha \hat{f}_{p(\alpha - h)}^* \hat{f}_{p^\prime (\alpha - h^\prime)}}{\sum_\alpha \hat{f}_{p (\alpha - h)}^* \hat{f}_{p (\alpha - h)}} \delta_{d d^\prime}.
\end{equation}
Notice that $\delta_{d d^\prime}$ in Eq.~(\ref{eq:window_apply_explicit}) means the window matrix is block diagonal, which we can take advantage of to make the computation more efficient. 

Similarly, using the definition of $\mbf{\Gamma}_{pdh}$ given in Eq.~(\ref{eq:gamma_def_apply}), we see that Eq.~(\ref{eq:b_hat_def}) becomes
\begin{equation} \label{eq:b_hat_simple}
    \hat{b}_{pdh} (\tau) = \frac{\sum_\alpha \hat{f}_{p (\alpha - h)}^* \hat{s}_{d \alpha} (\tau)}{\sum_\alpha \hat{f}_{p (\alpha - h)}^* \hat{f}_{p (\alpha - h)}}.
\end{equation}
Equation~(\ref{eq:b_hat_simple}) offers a simple interpretation of Eq.~(\ref{eq:b_hat_def}): to estimate the leakage parameters, we just need to perform a lag correction, with $\Delta \alpha = h$, along the RA axis between Dec $p$ of the foreground map and Dec $d$ of the signal map. Then we can compensate for the window in the estimates to obtain the unwindowed leakage estimates
\begin{equation} \label{eq:b_tilde}
    \tilde{b}_{pdh} (\tau) = \sum_{p^\prime, d^\prime, h^\prime} \mbf{W}_{(pdh)(p^\prime d^\prime h^\prime)}^+ \hat{b}_{p^\prime d^\prime h^\prime} (\tau),
\end{equation}
where $\mbf{W}_{(pdh)(p^\prime d^\prime h^\prime)}^+$ is the pseudo-inverse of the window matrix given by Eq.~(\ref{eq:window_apply_explicit}). Note that the window matrix might not be full rank. This happens because certain leakage parameters may be unconstrained from the estimation done with Eq.~(\ref{eq:b_hat_simple}), for example, when some declinations of the foreground map are masked out or if some declinations contain no or only a few weak foreground sources. However, if foreground sources are weak or not present at some declinations, they are unlikely to contaminate high delay maps, so we need not estimate such leakage.

Using the unwindowed leakage parameters, we can reconstruct the transfer matrix
\begin{equation}
    \mbf{\tilde{B}}(\tau) = \sum_{p,d,h} \tilde{b}_{pdh} (\tau) \mbf{\Gamma}_{pdh},
\end{equation}
and obtain the cleaned signal $\bm{\tilde{s}}$ by subtracting the foreground leakage from the estimated signal $\bm{\hat{s}}$:
\begin{equation} \label{eq:s_tilde}
    \bm{\tilde{s}} = \bm{\hat{s}} - \mbf{\tilde{B}} \bm{\hat{f}}.
\end{equation}
Note that similar to the general formalism in Section~\ref{subsec:form_rev}, the algorithm developed in this section does not subtract foregrounds exactly either. This is because in this case, we have assumed foreground leakage at high delays comes from the zero delay map only, while in reality foreground power from other low delays may contribute to high-delay foreground leakage as well. In addition, the foreground sources in the zero delay map are also convolved with CHIME's primary beam. The unremoved primary beam structures in the zero delay map may introduce second-order errors when we subtract beam-induced leakage at high delays. Finally, certain compromises have to be made when we implement the algorithm on top of the CHIME stacking pipeline, which we will examine in the following section.

\section{Implementation} \label{sec:app}

\subsection{CHIME stacking pipeline}

To remove foreground residuals in the CHIME stacking analysis, we need to implement HyFoReS on top of the CHIME stacking pipeline, which is described in detail in \citep{chime_stacking}. Here we provide a short summary of the main steps in the stacking pipeline.

The input data for the CHIME stacking pipeline is a sidereal stack, which is the visibility for all unique baselines as a function of local Earth-rotation angle at $\Delta \nu = 0.390625$~MHz spectral resolution averaged over sidereal days. Sidereal stack was produced through the CHIME real-time and daily processing pipelines using the raw data collected by the CHIME telescope over 102 nights. We refer readers to \citep{chime_overview} and~\citep{chime_stacking} for these details.

The sidereal stack then goes through the stacking pipeline, which includes the following steps: 1) point source subtraction, which estimates and subtracts signal of the four brightest point sources, namely Cygnus A, Cassiopeia A, Taurus A, and Virgo A, from the data; 2) frequency mask: mask out frequencies with RFI contamination, missing data, excessive noise, and frequencies near large gaps of missing data. This step removes a total of $48.6\%$ of the $587.5 - 800$ MHz band covered in the analysis; 3) map making which includes north-south beam forming and primary beam deconvolution (more details in Appendix~\ref{app:decon}), east-west beam forming, and a normalization procedure that preserves point source flux; 4) foreground filtering: apply the DAYENU delay filter described in Section~\ref{sec:motiv}; 5) outlier mask: remove any map pixel whose absolute value is greater than 6$\sigma$ (six times the standard deviation of the radiometric noise and persistent RFI); finally, 6) stacking: the foreground-filtered and outlier-masked map $m^q(\nu, \theta, \phi)$, where $q$, $\nu$, $\theta$, and $\phi$ stand for polarization, frequency, declination, and right ascension, respectively, is stacked at angular and spectral locations of the eBOSS catalog sources. These steps are summarized in blue boxes of Fig.~\ref{fig:pipeline}.

Converting the redshift of catalog objects to frequencies,
\begin{equation}
    \nu_{21\text{cm}} = \frac{1420.406 \text{ MHz}}{1 + z},
\end{equation}
the stacked signal, measured as a function of the frequency offset $\Delta \nu$, is simply
\begin{equation}
    d^q(\Delta \nu) = \langle m^q(\nu_s + \Delta \nu, \theta_s, \phi_s) \rangle_s,
\end{equation}
where $\langle ... \rangle_s$ represents a weighted average over all the catalog objects $s$, with the weights being the inverse variance of map pixels. In the CHIME stacking analysis, all three tracer catalogs (ELG, LRG, and QSO) were used. Among the three, the QSO catalog has the largest overlap in redshift with the CHIME frequency range chosen for that analysis (corresponding to $0.8 < z < 1.4$) and achieved the highest detection significance in the stacking analysis. For simplicity, we only perform stacking on the QSO catalog for this study.

\subsection{Implementing HyFoReS} \label{subsec:implement}

\begin{figure}[h]
\includegraphics[scale=0.40]{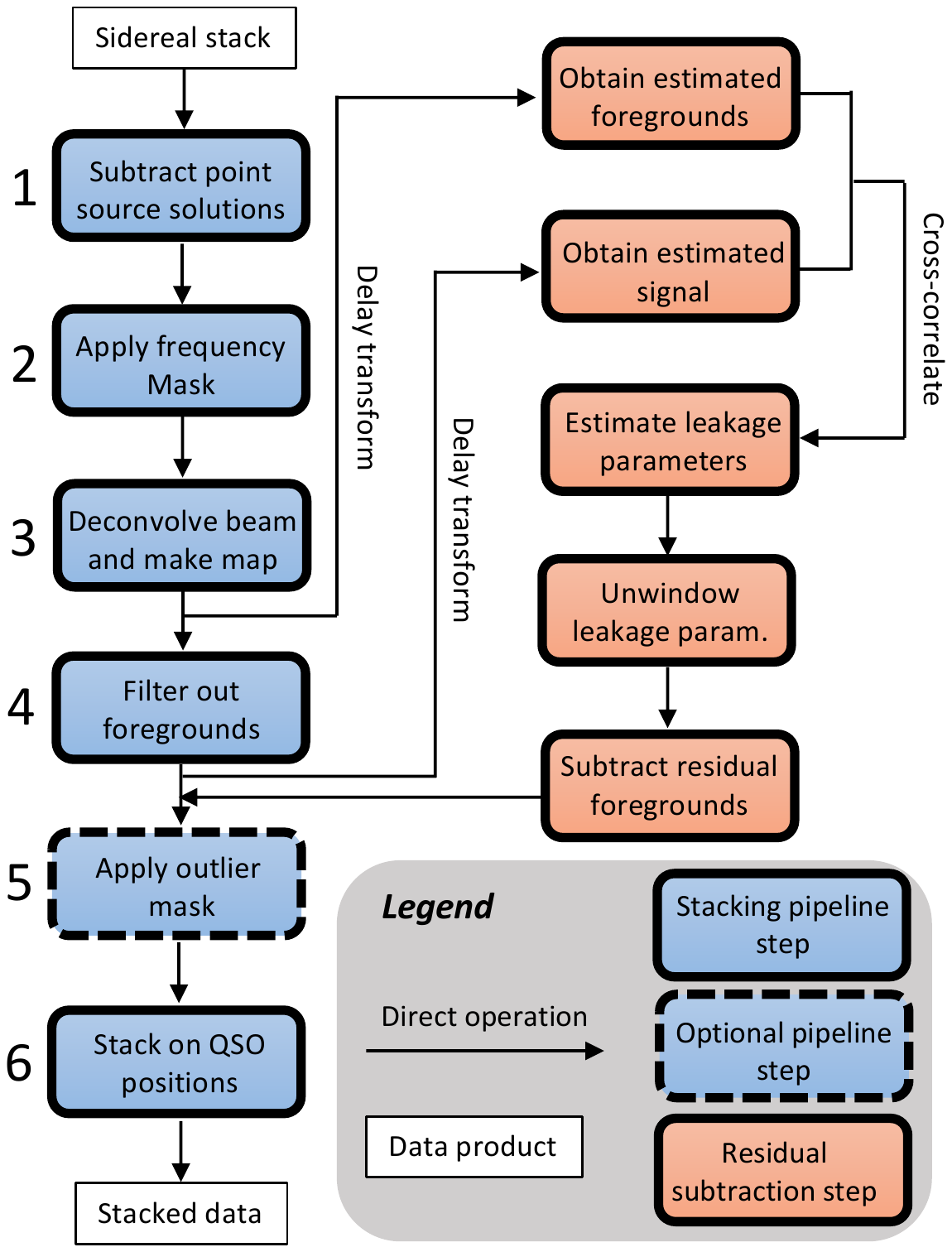}
\caption{\label{fig:pipeline} A schematic representation of the foreground residual subtraction algorithm, implemented on top of the CHIME stacking pipeline \citep{chime_stacking}. The CHIME stacking pipeline uses the sidereal stack (visibilities averaged over 102 nights) as the input, subtracts the four brightest point sources, applies a frequency mask for missing/noisy data and consistent RFI, deconvolves the primary beam and makes the map, removes foregrounds through a high-pass delay filter, applies an outlier mask that zeros out bright pixels from residual foregrounds, and finally stacks the map on the positions of eBOSS catalog objects. The resulting stacked data still contains residual foregrounds. We introduce the HyFoReS algorithm to the stacking pipeline to remove beam-induced foreground residuals. Estimated foregrounds and signal are obtained from the pre-filtered and post-filtered data, respectively. The algorithm then cross-correlates the two to draw out the foreground leakage parameters, which are used to subtract residual foregrounds from the map. The residual-subtracted map then propagates down the pipeline. In this analysis, we can turn the outlier mask on and off to test the effect of the subtraction algorithm alone at removing residual foregrounds as well as its combined effect with the outlier mask. }
\end{figure}

To implement HyFoReS, we need to modify the stacking pipeline and inject additional steps. This is summarized in Fig.~\ref{fig:pipeline}. As described in Section~\ref{sec:form}, HyFoReS requires estimated foregrounds and signal, which are obtained by applying linear filters on the data. As in Eqs.~(\ref{eq:f_hat_algo}) and (\ref{eq:s_hat_algo}), the estimated foregrounds and signal are sky maps at zero and high delays, respectively. Note that the sky map is measured as a function of frequency, i.e., a frequency spectrum. In principle, a direct delay transform on the map (Fourier transform performed along the frequency axis) would give us the delay spectrum. However, masked frequency channels in the map would mix power across delays via Fourier transform, causing low delay foregrounds to contaminate high delay signal. To mitigate this challenge, we use a Gibbs sampling technique to perform the delay transform, which is a statistical approach that can compensate for the lost frequency information in the data due to masking and avoid mixing of power in the obtained delay spectrum.  These details are provided in Appendix~\ref{app:dspec}.

We delay transform the sky map before the DAYENU delay filter is applied (between step 3 and 4 in the stacking pipeline) and extract the zero delay map as the estimated foregrounds. Similarly, the estimated signal is obtained by performing another delay transform on the map immediately after the DAYENU delay filter (between step 4 and 5 in the stacking pipeline). We mask out the regions on the maps that correspond to the four brightest point sources (Cygnus~A, Cassiopeia~A, Taurus~A, and Virgo~A) picked up by the far side lobes of the primary beam, three of which are visible as bright arches across the estimated signal map in the right panel of Fig.~\ref{fig:low_vs_high}. As described in the appendix of \citep{chime_stacking}, given a point source at its true declination $\theta^\prime$ and right ascension $\phi^\prime$, the far side lobes of the primary beam leave a U-shaped track around the source on the map, parametrized by an effective declination of the source $\theta^\prime_{\text{eff}}$ at each right ascension $\phi$:
\begin{equation}
    \theta^\prime_{\text{eff}}(\phi) = \arcsin(\cos{\Lambda}\sin{\theta^\prime} - \sin{\Lambda}\cos{\theta^\prime}\cos{(\phi - \phi^\prime)}) + \Lambda,
\end{equation}
where $\Lambda$ is CHIME's telescope latitude. The mask is constructed by zeroing out the regions between $\theta^\prime_{\text{eff}}(\phi) - 1.5^\circ$ and $\theta^\prime_{\text{eff}}(\phi) + 1.5^\circ$ at each right ascension $\phi$ of the data map for each of the four brightest point sources. This masking is to ensure the following steps are not contaminated by these four particular sources, which are located outside of the stacking analysis region but nonetheless leave significant residuals through far side lobes.   


We then cross-correlate the estimated foregrounds from $\tau = 0$ with estimated signal at $\tau > 0$ to obtain the low-to-high-delay foreground leakage parameters (Eq.~[\ref{eq:b_hat_simple}]). In the current implementation of the algorithm, we assume the intrinsic foregrounds are only at zero delay. We therefore estimate the leakage parameters $\hat{b}_{pdh} (\tau)$ for all $\tau > 0$, including the delays below the cutoff of the DAYENU delay foreground filter $\tau_{\text{cutoff}}$. This is because even though the DAYENU delay filter has already suppressed power below $\tau_{\text{cutoff}}$, it does not simply zero out all power below that cutoff. Estimating foreground leakage and performing foreground residual subtraction over all delays, including those below $\tau_{\text{cutoff}}$, can in principle reduce foregrounds missed by the DAYENU delay filter.

In addition to being a function of delay, the leakage parameter $\hat{b}_{pdh} (\tau)$ also depends on declination of the foreground map $p$, declination of the signal map $d$, and the relative RA difference between the two maps $h$. Technically, the Dec $d$ and RA difference $h$ should cover the entire signal map when we consider how a source from the foreground map leaks into signal maps, but this is computationally challenging. More importantly, the full range of $d$ and $h$ would make the parameter space too large for the leakage parameters $\hat{b}_{pdh} (\tau)$ to be constrained accurately from the available data. Modifications of our formalism may be able to resolve this issue, and we plan to investigate this in future work.

As a first study on removing beam-induced residuals using this type of algorithm, we aim to subtract leakage only within a narrow region around each foreground source. Foreground source leakage seen in Fig.~\ref{fig:low_vs_high} suggests that the main lobe of the point spread function is fairly narrow, especially in the declination direction. Therefore, we choose $-0.88^\circ < h < +0.88^\circ$, which covers 21 pixels along the RA direction, and $p -0.2^\circ < d < p +0.2^\circ$, which covers 5 pixels along the declination direction, forming a rectangular region centered around any foreground source located at Dec $p$. Then Eq.~(\ref{eq:b_hat_simple}) estimates the leakage parameters $\hat{b}_{pdh}$ within this rectangular region by averaging over the RA axis at each Dec $p$ of the foreground map. 

\begin{figure*}
\includegraphics[scale=0.75]{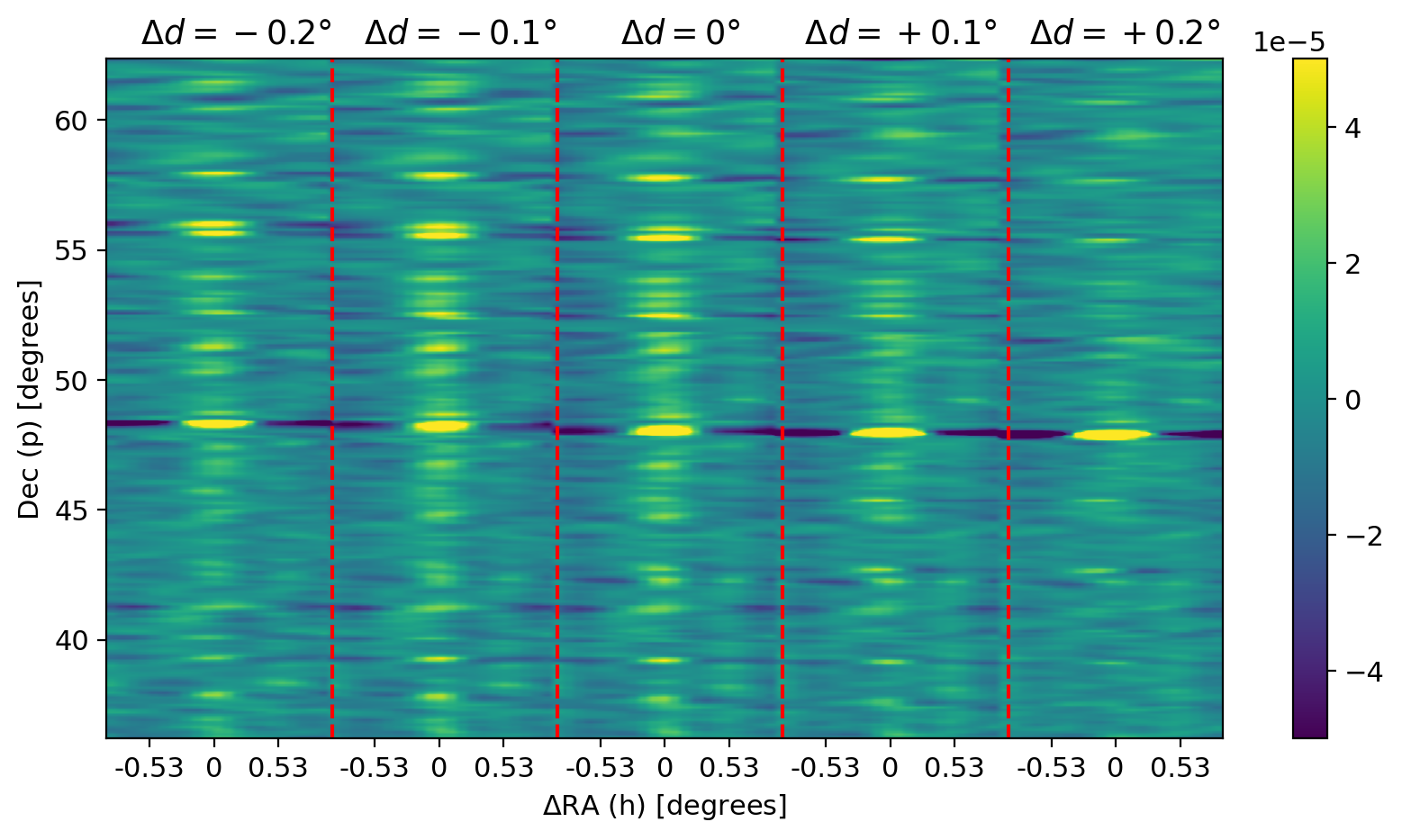}
\caption{\label{fig:leakage_par} Unitless leakage parameter $\hat{b}_{pdh}$ estimated by cross-correlating the zero-delay foreground map with the high delay signal map at $\tau = 329 \text{ ns}$. Only the XX polarization is shown. The $y-$axis represents the Dec $p$ of the foreground map. Each column, separated by dashed red lines, shows leakage parameters measured at Dec $d$ of the signal map with the relative Dec difference $\Delta d = d - p$ given at the top. The bottom $x-$axis shows the relative RA difference $h$, ranging from $-0.88^\circ$ to $+0.88^\circ$ for each column. Even though we limit the range of our cross-correlation, the leakage parameters are able to estimate the central region of the point spread function, which contributes the most to foreground leakage.
}
\end{figure*}

The estimated leakage parameters are shown in Fig.~\ref{fig:leakage_par} at $\tau = 329$ ns. Each column, separated by dashed red lines, represents a declination in the signal map near the foreground map Dec $p$, with the relative Dec differences shown at the top. Our choice of $h$ allows the leakage parameter estimation to cover the central peak of the point spread along the RA axis, where foreground residuals are most severe. Note that we detect foreground leakage that spreads in Dec and RA (i.e.\ $\hat{b}_{pdh}\neq 0$ at $\Delta d \neq 0$ and $h = \Delta{\rm RA} \neq 0$). The spread matches our expectations from foreground contamination seen on the right panel of Fig.~\ref{fig:low_vs_high} and is likely due to the chromatic beam that has not been deconvolved from the map. Figure~\ref{fig:leakage_par} also suggests that leakage of foreground power from $\tau = 0$ to $329\text{ ns}$ is roughly at the order of $10^{-4}$. Given that foregrounds are about $10^4$ to $10^5$ times brighter than the HI signal, we expect foreground residuals to dominate within the contaminated regions of the map.


We then compensate for the window in the leakage parameters using Eq.~(\ref{eq:b_tilde}) and finally subtract foreground residuals from the estimated signal using Eq.~(\ref{eq:s_tilde}). The cleaned signal would then go through step 5 and 6 of the stacking pipeline. Notice that both the outlier mask and HyFoReS mitigate foreground residuals, but they do so from two different perspectives. The outlier mask removes any foreground residual and artifact that exceeds the $6 \sigma$ threshold but leaves out fainter contamination. HyFoReS removes foreground residuals in the vicinity of any foreground source, which may include fainter residuals under the $6 \sigma$ threshold, but leaves out foregrounds seen in far side lobes and other artifacts that are not correlated with nearby foreground sources. In principle, combining both HyFoReS and the outlier mask can remove most foreground residuals, but this also risks higher signal loss (see Section \ref{subsec:signal_loss}). 

In this work, we take a three-step approach to study the effectiveness of applying HyFoReS to the CHIME stacking analysis. First, we skip the outlier mask and directly stack on the residual-subtracted map (namely after HyFoReS implemented on the map). This is to isolate the effect of HyFoReS on the data and demonstrate its ability of removing residual foreground. Next, we turn on the outlier mask and stack on the residual-subtracted and outlier-masked map, which is to test if the addition of HyFoReS improves the detection significance of the original CHIME stacking analysis. Finally, we repeat this process but with lower delay cutoffs for the DAYENU delay filter. This is to test whether HyFoReS allows a more relaxed foreground filter, i.e., a filter that preserves more information at delays below the cutoff $\tau_{\text{cutoff}}$ of the DAYENU delay filter in the original stacking analysis. We will present our results in Section~\ref{sec:res} following this three-step approach.


\subsection{Quantifying detection significance} \label{subsec:detect_sig}

To quantify the detection significance of 21 cm signal in the stacked data, we follow the procedure described in \citep{chime_stacking}, fitting a 21 cm signal-only template to the stacked data and determining the signal-to-noise ratio from the fit. The stacked data $d^q(\Delta \nu)$ contains noise which includes contributions from remaining foregrounds and RFI. Due to the effects of the foreground filter and residual foregrounds, noise is correlated between frequencies. To characterize the noise covariance, we use the eBOSS QSO ``random'' catalog \citep{eboss_rand_1, eboss_rand_2} , which approximates the selection function of the true catalog but is 40 times as dense. We sample a subset from the random catalog that contains the same number of sources as the true catalog and stack the data map on the subset to obtain a noise stack. We repeat this process 10,000 times. The noise covariance is then estimated from the covariance of the noise stacks.

To generate a signal template, we use the same simulations the CHIME stacking analysis implemented in \citep{chime_stacking}. In summary, a pair of HI and galaxy number density maps are generated from a Gaussian realization of an input matter power spectrum which includes nonlinear halo model predictions \citep{Mead_2021}. A mock QSO catalog is then drawn from the galaxy density map, while the HI density map is transformed into signal-only visibilities through the $m$-mode formalism \citep{mmode}. The signal-only visibilities are injected into the sidereal stack of the real data, which is then processed by the same stacking pipeline described in the previous subsection, with the only difference being that we now stack the injected map on the mock catalog instead of the true catalog. The resulting injected stack still contains noise from the real data, so we stack the real data map (without the injection) on the mock catalog to obtain the noise stack and subtract the noise stack from the injected stack to obtain the signal template.

Note that a minor but important difference between the original CHIME stacking analysis and this work is that in the original analysis, the signal template is made from the pure 21 cm simulation only. This is because the original CHIME stacking pipeline is mostly linear (nonlinear effects from the outlier mask were found to be at the few-percent level, subdominant to the statistical error on the measurement), so signal and foregrounds propagate independently through the pipeline. However, the implementation of HyFoReS in this work is a nonlinear operation on the map. Combining it with the outlier mask may cause foregrounds to affect the signal template in a nontrivial way, which is why we choose the injection approach to make the signal template.


The signal template is then fit to the stacked data with an overall amplitude $A$ and frequency offset $\Delta f$ as fitting parameters. The number of sigma detection (namely the signal-to-noise ratio) $N_A$ is estimated by dividing the best fit amplitude over the standard deviation of the posterior distribution of the fitting amplitude, which well approximates a Gaussian distribution. Note that the stacking analysis \citep{chime_stacking} used two additional methods to quantify the detection significance, i.e., the Bayes factor and likelihood-ratio test. All three methods gave consistent evaluations on the detection significance. For simplicity, we only use $N_A$ to evaluate detection significance in this work.


\section{results} \label{sec:res}

We present the results in three subsections.  In subsection~\ref{subsec:res_1}, we stack on the foreground-filtered and residual-subtracted map without implementing the outlier mask. Subsection~\ref{subsec:res_2} is done with the outlier mask but otherwise identical to the previous subsection. In subsection~\ref{subsec:res_3}, we repeat the analyses done in the previous two subsections but experiment with lower delay cutoffs for the high-pass foreground filter. To test whether HyFoReS improves the detection of 21 cm signal, we compute the signal-to-noise ratio $N_A$ for the procedure performed in each subsection and compare it with the signal-to-noise ratio of the same procedure but without HyFoReS.

\subsection{Applying HyFoReS alone} \label{subsec:res_1}

In this subsection, we test the ability of HyFoReS at removing beam-induced foreground residuals alone without the outlier mask. Recall that both HyFoReS and the outlier mask can remove some foreground residuals. To isolate the effect of the former, we propagate the sidereal stack through all steps shown in Fig.~\ref{fig:pipeline} except the outlier mask.

The effect of foreground residual subtraction can be seen directly through the delay power spectrum $P(\tau,\theta)$ of the map, which is defined as the variance of the delay spectrum at each delay and declination (see Appendix~\ref{app:dspec} for details). To give the delay power spectrum $P(\tau,\theta)$ the units of $(\text{Jy/beam})^2 \text{ MHz}$, we normalize it with
\begin{equation}
    P_{\text{norm}}(\tau,\theta) = \frac{\Delta \nu}{N_\nu^2} P(\tau,\theta),
\end{equation}
where $\Delta \nu = 212.5 \text{MHz}$ is the bandwidth of the CHIME stacking data, and $N_\nu = 544$ is the number of frequency channels. Note that the CHIME telescope feeds observe the 400 to 800 MHz band with 1024 frequency channels, but the CHIME stacking data was selected to only include 544 channels covering 587.5 to 800 MHz (corresponding to $z = 1.42 - 0.78$). This is to avoid persistent RFI bands and to maximize the overlap with eBOSS catalog objects in redshift. For the rest of the paper, we will refer to the normalized delay power spectrum simply as the delay power spectrum.

\begin{figure*}
\includegraphics[scale=0.22]{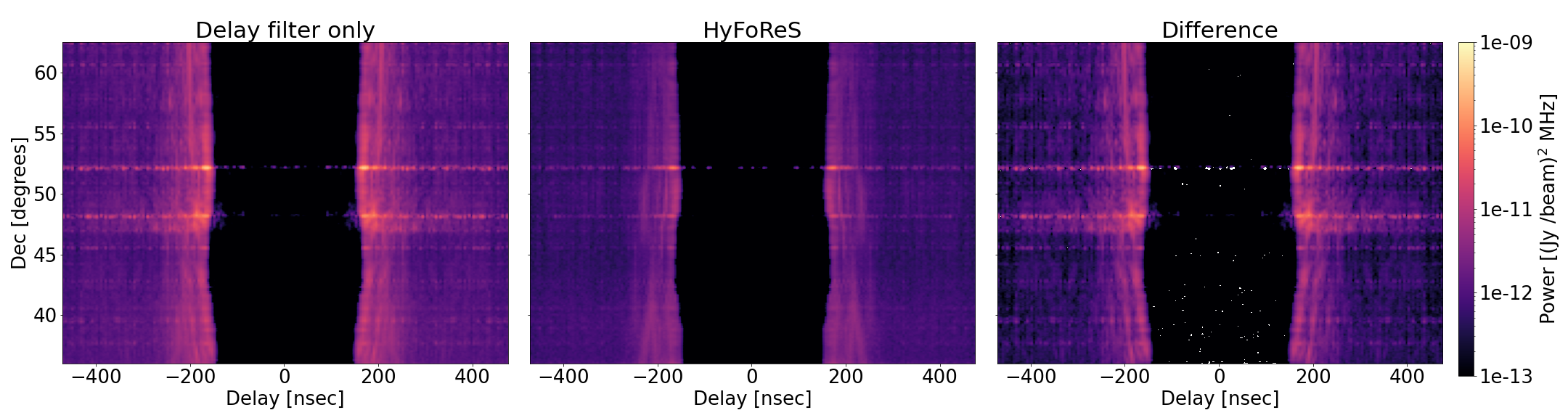}
\caption{\label{fig:dspec} Delay power spectrum, i.e. the variance of the delay space map along the RA axis, before and after HyFoReS and the difference. The data is sensitive out to 1250 ns but only the portion within 500 ns is shown. Left panel: Delay power spectrum of the delay-filtered map before HyFoReS is applied. Power between roughly -200 ns and 200 ns is suppressed by the delay foreground filter, but foreground residuals leak beyond the central cut region. Middle panel: delay power spectrum after HyFoReS is applied. Right panel: difference between the left and middle panel. HyFoReS suppresses power from point sources over all delays (seen as horizontal stripes on the plot at constant declinations) and reduces diffuse foreground power around 200 ns.}
\end{figure*}

The delay power spectra of the map before and after HyFoReS as well as their difference are shown in Fig.~\ref{fig:dspec}. Note that even though the data is sensitive out to 1250 ns, only the portion within 500 ns is shown because the power beyond 500 ns is dominated by the radiometric noise. The central gaps in these plots are due to the high-pass delay filter, which has removed most power under 200 ns (see Fig.~\ref{fig:cutoffs} for details of the relaxed cutoff). Two particular features of the delay power spectrum stand out. First, a few bright horizontal stripes can be found to stretch over all delays at some declinations. They correspond to point sources seen through the central peak of the beam. Second, there are bright vertical diffuse structures located just outside of the central cut region. They are likely due to interference between the primary light path and multiple reflections within the telescope as well as point sources picked up by the side lobes of the beam. We observe that HyFoReS is able to remove power from both structures, so the overall level of foreground contamination in the map is reduced after the algorithm.


\begin{figure*}
\includegraphics[scale=0.29]{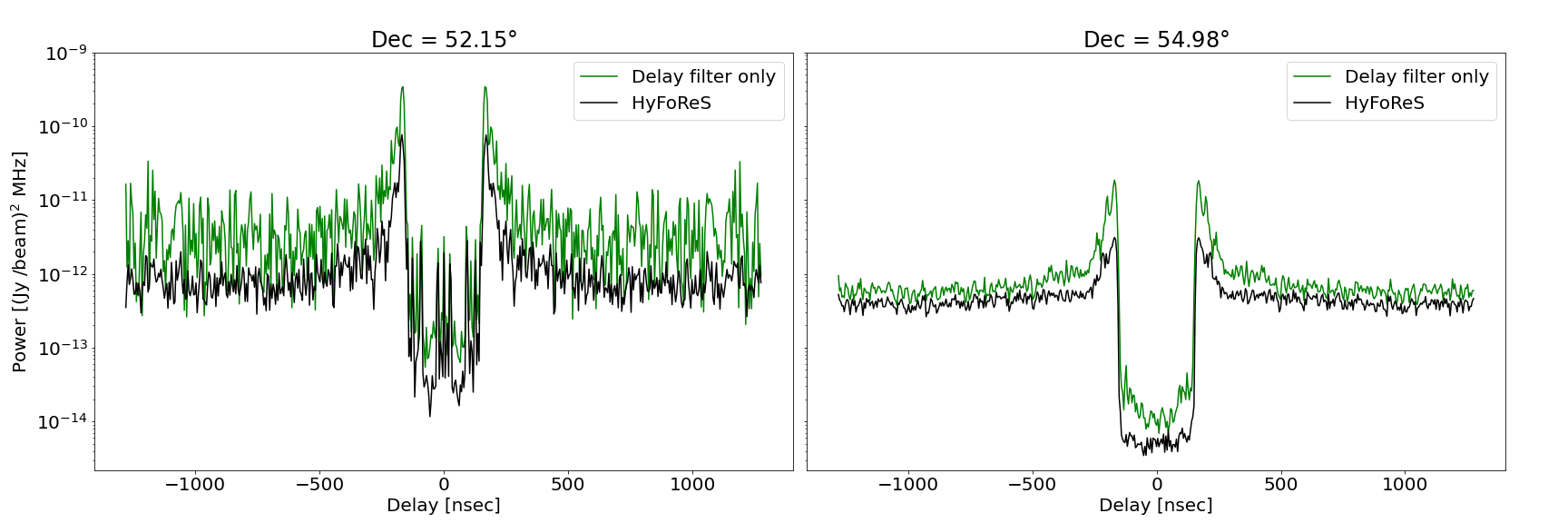}
\caption{\label{fig:dspec_2d} Delay power spectrum at two constant declinations before and after HyFoReS. Left panel: delay power spectrum at Dec $52.15^\circ$, which contains a bright foreground source. The power after HyFoReS (black curve) is nearly one order of magnitude lower than the power from the delay filtered map (green curve). Right panel: delay power spectrum at Dec $54.98^\circ$, where strong foreground sources are absent. HyFoReS only removes about 20\% of the original power. By removing more power from highly contaminated declinations and less from the cleaner declinations, the algorithm is able to reduce foregrounds while controlling signal loss. (Note that we have implemented HyFoReS over all delays, including those below $\tau_{\text{cutoff}}$ set by the DAYENU delay filter, so some power is also reduced at $|\tau| < 200$ ns.)  
}
\end{figure*}

We further compare the left and middle panels of Fig.~\ref{fig:dspec} by taking slices at Dec $= 52.15^\circ$, which contains a bright point source, and at Dec $= 54.98^\circ$, which is relatively foreground quiet. The slices are plotted in Fig.~\ref{fig:dspec_2d}. At Dec $= 52.15^\circ$, the power is reduced by nearly one order of magnitude after HyFoReS, while the power reduction is only about $20 \%$ at Dec $= 54.98^\circ$. This shows that HyFoReS reduces power more aggressively where foreground contamination is severe and preserves more power where foreground sources are absent. This is expected, since HyFoReS by design removes power from the signal map only when it is correlated with the foregrounds, although spurious correlations and noise in the correlation estimate can nonetheless lead to signal loss (see Section~\ref{subsec:signal_loss}).

\begin{figure*}
\includegraphics[scale=0.20]{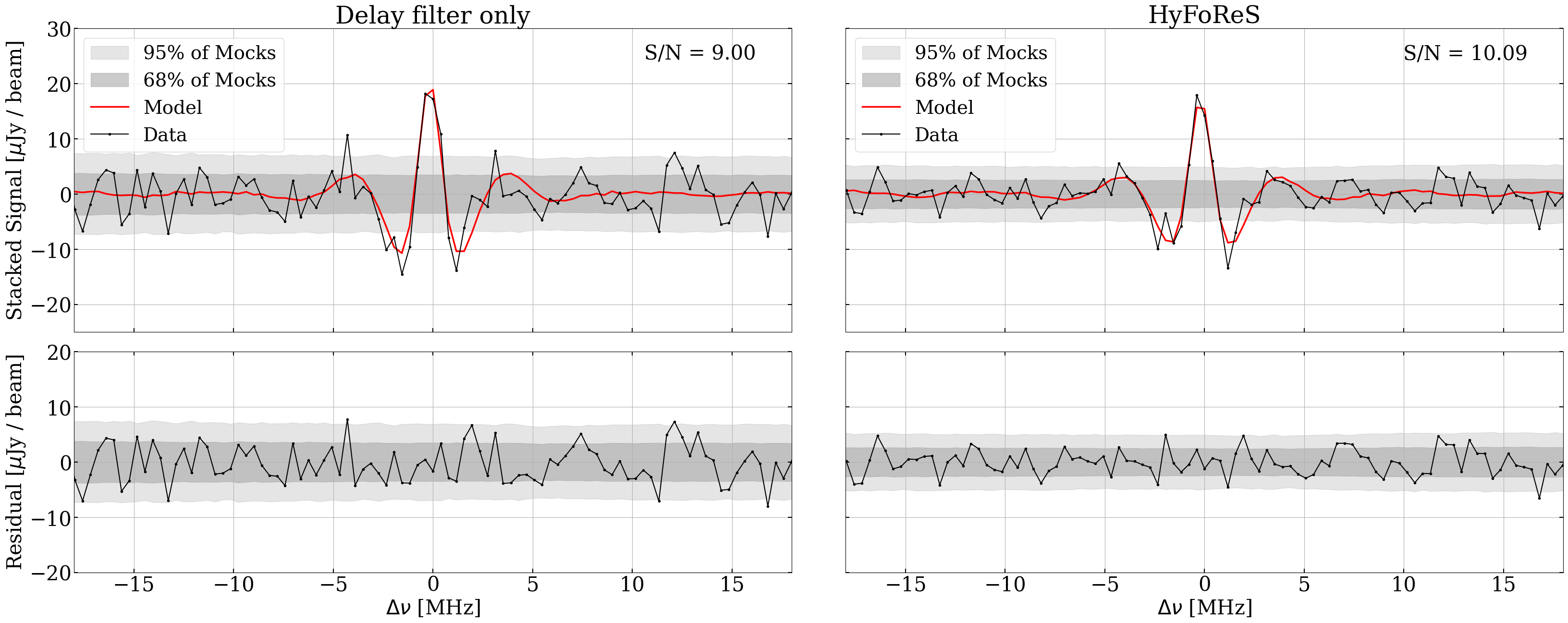}
\caption{\label{fig:fit_relax} Top: stacked signal as a function of frequency offset with the delay filter only (left) and with HyFoReS implemented after the delay filter (right). The stacked data are shown as black curves along with their best-fit models in red. The dark and light gray bands correspond to the central $68\%$ and $95\%$ of the values observed by stacking the data on 10,000 mock catalogs. The signal-to-noise ratios are quoted at the upper right corner of each panel, indicating the subtraction algorithm improves the signal-to-noise by more than $10\%$. Bottom: residuals obtained by subtracting the best-fit model from the data with the same noise bands shown as the top. The fitting residuals are consistent with the estimated noise level in both panels.}
\end{figure*}

Following the procedure outlined in Section~\ref{subsec:detect_sig}, we then fit a signal template to the stacked data and compute the signal-to-noise ratio $N_A$. The fitting results with the delay foreground filter only and with HyFoReS are shown in Fig.~\ref{fig:fit_relax}. Compared the delay filter only result, the amplitude of the best fit signal model (red curve) drops slightly after HyFoReS, indicating some signal loss, but the noise level (as indicated by the grey noise bands) also decreases. This leads to an increase of the signal-to-noise ratio by more than $10\%$. The fitting results show that the subtraction algorithm can reduce foreground residuals and improve signal detection when it is implemented without the outlier mask.

\subsection{Combining HyFoReS with outlier mask} \label{subsec:res_2}

\begin{figure*}
\includegraphics[scale=0.20]{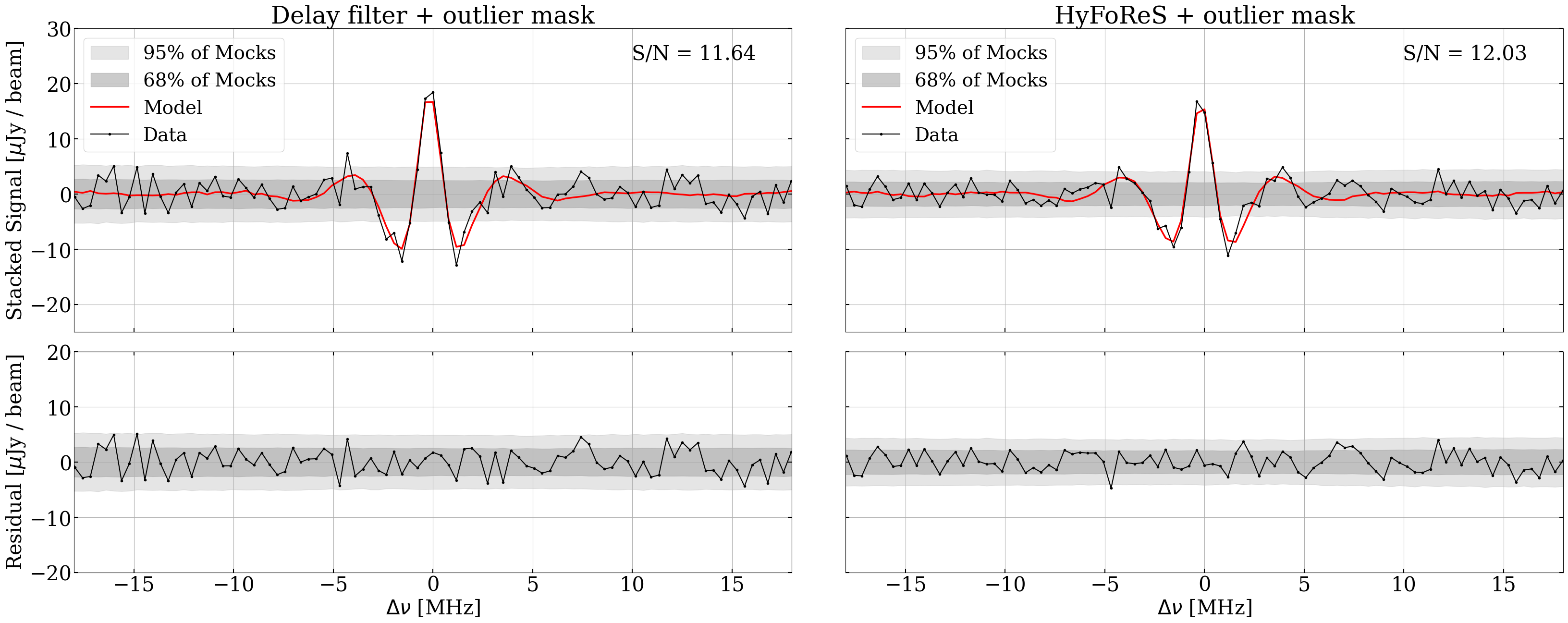}
\caption{\label{fig:fit_20} Same as the previous figure but with the outlier mask applied before stacking. Left: the outlier mask is applied to the delay-filtered map without HyFoReS. Right: HyFoReS is applied to the delay-filtered data, followed by the outlier mask. When the outlier mask is used, applying HyFoReS increases the signal-to-noise by roughly $4\%$.}
\end{figure*}

Now we study the combined effect of HyFoReS and outlier mask. We repeat the same procedure in the previous subsection but now include the outlier mask in the pipeline. Namely, we apply HyFoReS to the post-delay-filtered map, followed by the outlier mask. We then stack the map on eBOSS QSOs and fit a corresponding signal model. The result is shown on the right panel of Fig~\ref{fig:fit_20}. For comparison, we also fit the data with the delay foreground filter and outlier mask applied but without HyFoReS (i.e., identical to the CHIME stacking analysis), shown on the left panel of Fig~\ref{fig:fit_20}. Applying HyFoReS once again decreases both the stacked signal amplitude and noise band, which leads to an increase in the signal-to-noise by about $4\%$. 

The signal-to-noise does not increase as much as the previous case because the outlier mask has already removed some foreground residuals that HyFoReS can in principle remove as well. Recall that the outlier mask and HyFoReS target different kinds of residuals: the former removes contamination 6 times brighter than the standard deviation of map pixels while the later subtracts residuals correlated with the zero-delay foreground map. However, there is overlap between the two kinds, making the improvement after applying HyFoReS less significant.

Comparing the left panel of Fig~\ref{fig:fit_20} with the right panel of Fig~\ref{fig:fit_relax}, it is evident that applying the outlier mask alone reaches a higher signal-to-noise than using HyFoReS alone. There are two reasons. First, the current implementation of the algorithm cannot remove foreground residuals due to side lobes as well as artifacts such as RFI contamination.  Second, the 6$\sigma$ mask only affects a small portion of the map and therefore preserves more signal, while the foreground residual subtraction is done over every pixel of the map which may lead to more signal loss (see Section~\ref{subsec:signal_loss}). However, we notice that the highest detection significance is achieved when we combine both HyFoReS and the outlier mask as indicted by the right panel of Fig~\ref{fig:fit_20}.

\subsection{Relaxing the delay filter cutoff} \label{subsec:res_3}

Lastly, we study whether HyFoReS would allow us to lower the delay cutoff of the linear foreground filter. As discussed in Section~\ref{sec:motiv}, the delay cutoff of the original CHIME stacking analysis (referred to as the ``relaxed cutoff") excludes all BAO information from the filtered data and restricts the accessible physical scales in the nonlinear regime. There is an incentive to lower the delay cutoff to recover more large-scale information from the data while controlling foreground contamination. We repeat the procedure done in the previous two subsections but with the delay cutoff set to be 20~ns and 40~ns below the relaxed cutoff used in the CHIME stacking analysis (see Fig.~\ref{fig:cutoffs}).

The results are summarized in Table ~\ref{tab:Na}. For each of the three delay cutoffs (relaxed, relaxed - 20ns, and relaxed - 40ns), we stack on four sets of data. Each set of data goes through the same stacking pipeline summarized in Fig.~\ref{fig:pipeline} but differs in whether HyFoReS and/or the outlier mask is applied. For simplicity, we only show the signal-to-noise ratios from fitting a signal template on each stack. Note that results from the previous two subsections are also included in the column corresponding to the relaxed delay cutoff.

We observe that the highest signal-to-noise ratio is achieved when using both HyFoReS and outlier mask with the relaxed-20ns delay cutoff, which has a 0.6$\sigma$ improvement from $N_A = 11.6\sigma$ in the original stacking analysis. The improvement is modest but shows the potential HyFoReS has at removing foreground residuals and relaxing the delay filter cutoff. Note that the detection significance drops, however, as we further decrease the delay cutoff. This is because at lower delays, there are substantially more foreground residuals (such as far sidelobes from bright point sources) and artifacts (most likely residual RFI) that neither the subtraction algorithm or outlier mask is designed to remove effectively at the current stage.


In addition, we observe that HyFoReS alone consistently improves the signal-to-noise by 10 to 20$\%$ compared with the data stacked without the algorithm or outlier mask. This shows that HyFoReS is successful at removing foreground residuals and preserving signal in the sense that it improves the detection significance in each case. On the other hand, the outlier mask outperforms HyFoReS due to the fact that the outlier mask targets broader types of systematics and preserves more signal at high delays. Last, combining both subtraction algorithm and outlier mask achieves the highest detection significance with all three delay cutoffs, although the improvement is small compared with using the outlier mask alone. A full interpretation of the results will be given in Section~\ref{subsec:signal_loss}.

The results overall show that even though the subtraction algorithm does not improve the CHIME stacking analysis substantially at its current stage, it is nonetheless a successful proof-of-concept for removing beam-induced foreground residuals to improve detection significance. We will discuss how HyFoReS can be improved in Section~\ref{subsec:improve} and potentially applied to other 21 cm intensity mapping experiments.

\begin{table*}[]
    \centering
    \caption{Signal-to-noise ratios from fitting a signal template on the stacked data, which are processed with various delay cutoffs and combinations of HyFoReS and outlier mask. Three cutoffs for the delay foreground filter are used: the relaxed cutoff from the original CHIME stacking analysis and cutoffs 20 ns and 40 ns below the relaxed cutoff (shown in Fig.~\ref{fig:cutoffs}). For each cutoff, we process the data four times, each with a different combination of whether using HyFoReS and outlier mask. HyFoReS alone (second row) improves detection significance by $10 \sim 20\%$ compared with not using any foreground residual mitigation method (first row). Using the outlier mask alone (third row) increases the signal-to-noise ratios even more, but the highest signal-to-noise ratios are achieved by combining both (fourth row). This shows that foreground residual mitigation methods relax the requirement on the delay cutoff, allowing the highest detection to be reached with the relaxed-20ns cutoff.}
    \begin{ruledtabular}
    \begin{tabular}{cccc}
     Procedure & Relaxed delay cutoff & Relaxed - 20ns & Relaxed - 40ns  \\
     \hline
     No HyFoReS, no mask & 9.00$\sigma$ & 8.92$\sigma$ & 7.40$\sigma$ \\
     With HyFoReS, no mask & 10.09$\sigma$ & 10.88$\sigma$ & 9.15$\sigma$ \\
     No HyFoReS, with mask & 11.64$\sigma$ & 11.66$\sigma$ & 10.36$\sigma$ \\
     With HyFoReS, with mask & 12.03$\sigma$ & 12.22$\sigma$ & 10.41$\sigma$ \\
    \end{tabular}
    \label{tab:Na}
    \end{ruledtabular}
\end{table*}

\section{Discussion} \label{sec:dis}

\subsection{Assumptions of the algorithm} \label{subsec:assump}

The algorithm is an application of the previously developed HyFoReS formalism, which relies on using a linear foreground filter to obtain foreground and signal estimates first and then cross-correlates the two to subtract foreground residuals from the latter. The main assumption of the algorithm is that contamination in the estimated signal is correlated with low-delay foregrounds. This is true in general, but contamination could also arise from unmasked RFI and excessive noise at particular frequencies and declinations and from other unknown origins. These types of contamination are indeed visible in the signal maps and can cause spurious correlation with the low-delay foregrounds, degrading the estimation of foreground leakage.  

Two particular choices in the current implementation require additional assumptions that are not always valid in the data. First, for simplicity, we use the zero delay map as the estimated foreground. Foregrounds have smooth spectral structures and therefore concentrate most power at the zero delay, but more than the zero-delay mode (namely the mean mode) is needed to fully capture foregrounds. It is possible that some foreground residuals in the estimate signal are unaccounted for by correlating only with the zero delay map.

Second, limited by computational resources and the amount of information available in the data, we restrict the region of correlation between the estimated foreground and signal maps to be within a box of $1.76^\circ$ by $0.4^\circ$ along the direction of RA and Dec, respectively (more discussion on the choice of the correlation size in Section~\ref{subsec:cor_size}). Although this choice covers the central region of the point spread function (see Fig.~\ref{fig:leakage_par}), side lobes of the CHIME's beam extends beyond this area \citep{chime_stacking}. Therefore, the current implementation of the algorithm cannot subtract foreground residuals due to side lobes.

More fundamentally, the correlation between foreground and signal estimates captures the two-point statistics of information shared by the two maps, which works well for Gaussian distributed sources but cannot capture all the information if the underlying field is non-Gaussian. In this analysis, foreground residuals mostly arise from point sources, which are Poisson distributed. The result is that a few excessively bright point sources can dominate other sources at a particular declination, making the correlation insensitive to the rest of the sources at that declination. Those excessively bright point sources can leak outside the correlation region through side lobes, further degrading the estimation of leakage parameters and residual subtraction over the whole map.

Lastly, HyFoReS only subtracts foregrounds up to the first order in perturbations. Due to the enormous foreground-to-signal ratio, foreground leakage from second-order perturbations can still overwhelm the signal if the systematic errors are large. This is usually remedied by an initial step of calibration and telescope characterization, so errors in the calibration and instrument response models are small. Even though the main lobe of CHIME's primary beam was calibrated in the stacking analysis, Appendix~\ref{app:decon} suggests that the deconvolution was unsuccessful, leaving significant beam effects in the map. The subtraction algorithm can still remove some beam-induced foreground residuals, but residuals from second-order beam perturbations limit the effectiveness of the algorithm.

\subsection{Choosing the correlation size} \label{subsec:cor_size}

When estimating the leakage parameters in Eq.~(\ref{eq:b_hat_simple}), for any pixel at Dec $p$ and RA $\alpha$ in the estimated foreground map, we correlate it only with nearby pixels in high delay maps with Dec $d$ and RA $\alpha^\prime$ satisfying $p - 0.2^\circ < d < p + 0.2^\circ$ and $-0.88^\circ < h < 0.88^\circ$ ($\alpha^\prime = \alpha + h$), respectively. Note that we have assumed stationarity in the RA direction due to the m-mode symmetry of our telescope response \citep{first_mmode, mmode}.

The correlation size ideally should cover all the areas affected by the whole primary beam, including side lobes, but this is not practical since larger correlation size means more parameters to estimate, which in turn requires more information from the data for the parameters to be accurately constrained. To ensure proper estimation of the leakage parameters, we require the number of pixels in the map to be one order of magnitude larger than the number of leakage parameters. The current choice of the correlation size corresponds to 105 leakage parameters at each Dec and delay. The correlation is performed along the RA axis which contains about 1,700 pixels per Dec and delay, thus meeting our requirement.

While smaller size of the parameter space reduces noise in estimation, we want the correlation size to be large enough to cover the central peak of the beam, where foreground leakage from point sources is most severe. Figure~\ref{fig:leakage_par} shows that the current choice of the correlation size captures the central peak of the beam. However, the CHIME stacking analysis suggests that the side lobes of the primary beam can still reach a significant fraction of its central peak in the frequency domain \citep{chime_stacking}. This means that to improve the algorithm, we need to expand the correlation region to account for side lobes around foreground sources, which is discussed in Section~\ref{subsec:improve}.

\subsection{Signal loss} \label{subsec:signal_loss}

\begin{figure*}
\includegraphics[scale=0.25]{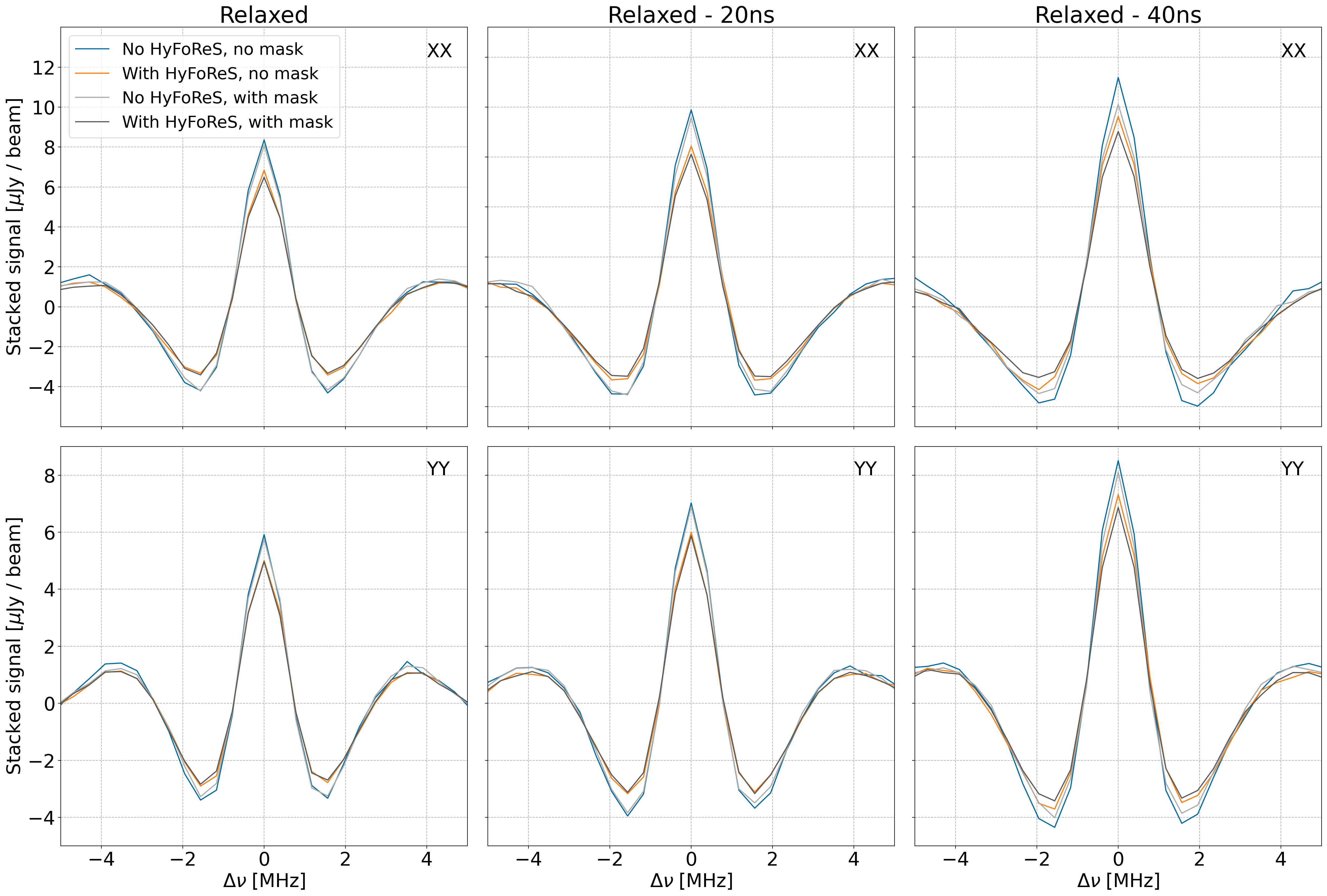}
\caption{\label{fig:sig_temp} Signal templates by stacking on the real data with signal injection. Left, middle, and right columns correspond to using the relaxed, relaxed-20ns, and relaxed-40ns cutoffs, respectively, for the delay foreground filter. The XX and YY polarization signal models are shown in the top and bottom panels, respectively. For the XX polarization, fractional signal loss due to HyFoReS alone is about $18\%$ regardless of which delay cutoff is used, while the signal loss due to the outlier mask increases from $3\%$ to $12\%$ as we lower the delay cutoff. Signal loss is greatest, reaching over $20\%$, when combining both HyFoReS and outlier mask. The fractional signal loss of the YY polarization is lower overall compared to the XX polarization but shows otherwise similar behavior to those of the XX polarization.}
\end{figure*}

To understand the effect of HyFoReS on the signal, we need to investigate signal loss after applying the algorithm. In this study, we quantify signal loss by comparing the amplitudes of the signal templates (used for fitting the HI signal of the stacked map) before and after HyFoReS. The signal templates are obtained by injecting realistic 21 cm simulation into the real data, with details described in Section~\ref{subsec:detect_sig}. Figure~\ref{fig:sig_temp} shows the signal templates. The three columns correspond to the three delay cutoffs implemented in Section~\ref{subsec:res_3}. The first and second row shows the XX and YY polarization signal templates, respectively. Four types of templates are shown in each panel, corresponding to the four foreground residual mitigation procedures shown in the first column of Table~\ref{tab:Na}, namely no foreground residual mitigation, HyFoReS only, outlier mask only, both HyFoReS and outlier mask.


We observe that HyFoReS and outlier mask both attenuate the signal compared with the template without either step. The fractional drop in the signal from HyFoReS alone stays roughly constant at $18\%$ for the XX polarization and $14\%$ for the YY polarization with all three delay cutoffs, as estimated through the data injection method described in Section \ref{subsec:detect_sig}. In comparison, the fractional signal loss due to the outlier mask alone is overall lower than that from HyFoReS but increases from $3\%$ using the relaxed cutoff to $12\%$ with the relaxed-40ns cutoff in the XX polarization. Signal loss in the YY polarization due to the outlier mask is smaller but also increases as the delay cutoff drops. Recall that the outlier mask zeros out pixels above 6 times the standard deviation of the radiometric noise. In the map filtered using the relaxed cutoff, we expect only a small fraction of pixels meet this criterion. On the other hand, HyFoReS can in principle remove power from every pixel in the map, so it is more susceptible to signal loss in comparison. However, as we relax the delay cutoff, more foreground residuals remain in the map, so more pixels stand above the $6\sigma$ threshold as well, causing more signal loss from the outlier mask.

Gathering evidence from the previous subsections, we now give a full interpretation of the results shown in Table~\ref{tab:Na}. HyFoReS improves the detection significance but is not as effective as the outlier mask when implemented alone because a few assumptions underlying the current implementation of HyFoReS are broken by the data (see Section~\ref{subsec:assump}). In addition, the outlier mask has the advantages of having less signal loss and targeting broader types of foreground residuals. Combining the outlier mask with HyFoReS further improves detection significance because HyFoReS cleans residuals from fainter foreground sources that are below the outlier mask's threshold. This improvement is mild due to the additional signal loss introduced by HyFoReS. 

As we further relax the delay cutoff, the data contains more signal but only to be overwhelmed by more foreground residuals from the lower delays. This leads to a decrease in signal-to-noise when neither HyFoReS or outlier mask is applied (see the first row of Table~\ref{tab:Na}). When either HyFoReS or outlier mask is applied, the additional amount of signal from the relaxed-20ns delay cutoff raises the detection significance slightly compared with the case of using the relaxed cutoff, with the highest detection significance achieved by combining both HyFoReS and outlier mask. However, the signal-to-noise turns over and decreases again as we push the cutoff to 40 ns below the relaxed cutoff because signal loss from the outlier mask further increases. In addition, more point sources picked up by side lobes and other artifacts make HyFoReS less effective. To improve the detection significance at lower delay, we may need to select a more generous threshold for the outlier mask (although risking more signal loss) and extend the correlation size of HyFoReS to account for side lobes.

\subsection{Potential of relaxing the delay cutoff}

\begin{figure*}
\includegraphics[scale=0.3]{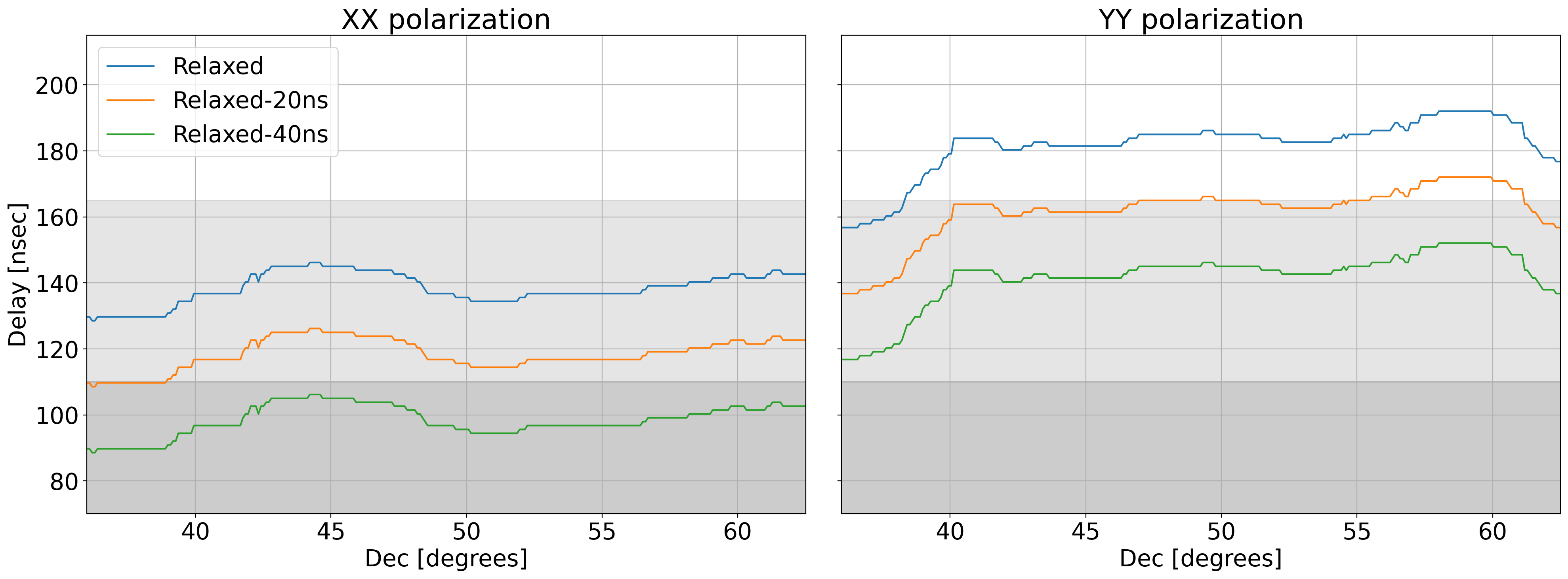}
\caption{\label{fig:cutoffs} Dec-and-polarization-dependent delay cutoffs used in this work. The light and dark gray bands correspond to regions where the data have sensitivity to the third and second peaks of BAO, respectively. The relaxed delay cutoff used in the original CHIME stacking analysis has nearly zero sensitivity to the third peak of BAO in the YY polarization. The relaxed-20ns cutoff captures the third peak over most of the Dec ranges in both polarizations, while full sensitivity to the third peak is only achieved with the relaxed-40ns cutoff.}
\end{figure*}

In Section~\ref{sec:res}, we first apply the relaxed cutoff used by the original CHIME stacking analysis for the delay foreground filter. The relaxed cutoff was selected 75 ns below the minimum delays where the measured power spectrum from data is less than 3 times the power spectrum of the expected thermal noise \citep{chime_stacking}. This cutoff is declination-dependent but has a characteristic value of roughly 150 ns for the XX polarization and 200 ns for the YY polarization across the declination ranges covered by eBOSS catalogs. 

There is a one-to-one correspondence between the physical scales along the line of sight (namely the comoving number $k_\parallel$) and the delay values, with the minimum accessible $k_\parallel$ in the data determined by the delay cutoff. In particular, the BAO peaks in Fourier space are located at multiples of the wave number $k_{\text{BAO}} \approx 0.064 \text{ } h^{-1} \text{Mpc}$. At 600 MHz ($z \approx 1.4$ for 21 cm signal), the first three maxima of BAO wiggles correspond to delays of 55 ns, 110 ns, and 165 ns, respectively. Therefore, the relaxed delay cutoff excludes most BAO information from the data.
     
On the other hand, the third peak of BAO is mostly covered by the relaxed-20ns cutoff and completely covered by the relaxed-40ns cutoff (see Fig.~\ref{fig:cutoffs}). Without using any foreground residual mitigation techniques, detection significance drops at these lower delays (see second and third rows of Table~[\ref{tab:Na}]).  However, we reach the highest detection significance of $12.22\sigma$ when applying both HyFoReS and outlier mask with the relaxed-20ns cutoff. The detection significance is lower using the relaxed-40ns cutoff but still reaches $10 \sigma$ either with the outlier mask alone or when combined with HyFoReS. Note that this improvement mostly comes from the outlier mask rather than HyFoReS, but we have nonetheless found evidence that foreground residual mitigation techniques allow the study of BAO using real 21 cm survey data, even though more careful analysis on signal loss and nonlinear effects may be needed before BAO information can be extracted from 21 cm data.

\subsection{Future improvements} \label{subsec:improve}

It is possible to mitigate all of the limitations mentioned in Section~\ref{subsec:assump}. First, it is possible to increase the correlation size without degrading the quality of parameter estimation if the data include a wider range of RAs or have been integrated for a longer period of time to lower the noise level. In addition, the shape of the correlation region need not be rectangular. The formalism developed in Section~\ref{sec:form} in fact allows the leakage parameters $b_{dph}(\tau)$ to cover an arbitrary region on the map at any Dec $d$. For example, CHIME's far side lobes are U-shaped arches, with their centers located around point sources and extending to higher declinations. We could adjust the correlation region to cover this U-shaped region to subtract residuals from far side lobes. For a given data set, it should be possible to determine the optimal size and shape of the correlation region to achieve the best signal-to-noise after residual subtraction, which we will leave for future studies.

The HyFoReS algorithm will benefit from additional masking that better targets RFI, excessive noise, and other artifacts in the data. In this study, we have found that masking out the far side lobes from the four brightest point sources on the sky improves the subtraction algorithm (masking out the four brightest point sources and their side lobes alone without implementing HyFoReS does not improve signal-to-noise significantly). Recall that the four brightest point sources are not located within the stacking map region, which means their residuals cannot be targeted or cleaned by HyFoReS. Similarly, masking out any significant contamination that is not correlated with the low-delay foregrounds will likely improve the results. Masking out some additional strong point sources inside the map can also potentially improve the correlation results, since this may alleviate the problem of having a few point sources dominating over all the other sources at their declinations.

Lastly, a better deconvolution procedure should significantly decrease the level of beam perturbations in the maps, which helps HyFoReS by decreasing the second order perturbations that our algorithm is unable to subtract. Alternatively, with simple modifications of the algorithm, we could accommodate to use an external point-source-only radio sky survey as the estimated foregrounds. In that case, the estimated foreground map will be free of CHIME's beam effects, potentially making the estimation of beam perturbations at high-delay signal maps easier.

\subsection{Applicability to other 21cm experiments}

In this study, we applied the HyFoReS formalism to remove beam-induced foreground residuals in the CHIME stacking data. The formalism reviewed in subsection~\ref{subsec:form_rev} is general and has a wide range of applicability to residuals induced by other types of systematics such as complex antenna gains \citep{haochen_first_paper}. The algorithm in general requires a linear foreground filter to produce foreground and signal estimates. A second requirement is symmetries in the data over which we can cross-correlate the foreground and signal estimates to draw out foreground residuals. 

In the CHIME stacking analysis, the linear foreground filter was the DAYENU delay filter, but the algorithm can in fact work with any linear filter, such as the KL filter. For CHIME, we utilize the m-mode symmetry, which assumes RA stationarity in the instrument response such that foreground leakage has the same behavior for all foreground sources at a particular declination. This allows cross-correlation along the RA axis to estimate the leakage parameters. Similar symmetries likely exist in other experiments. For example, for an interferometer consisting of dish arrays, the time-dependent perturbations can be averaged out when observing a particular sky patch over an extended period of time. The leakage parameters can then be estimated over different sky patches to allow foreground residual subtraction using HyFoReS.

\section{Conclusions} \label{sec:conclu}

We have applied the hybrid foreground residual subtraction (HyFoReS) formalism to the CHIME stacking data in order to remove beam-induced foreground residuals. The original CHIME stacking analysis implements a high-pass delay filter to remove foregrounds, which produces a foreground and signal estimate from the data, but the latter is contaminated by foreground residuals due to CHIME's chromatic primary beam. Our algorithm cross-correlates the foreground and signal estimates to draw out the foreground residuals and subtract them from the latter. 

The original CHIME stacking pipeline also implements an outlier mask to suppress foreground residuals after the delay filter. Without applying the outlier mask, we find HyFoReS can improve the signal detection by $10\%$ to $20\%$, compared with using the delay filter alone. Combining HyFoReS with the outlier mask further improves the signal-to-noise ratio, although the improvements are marginal compared to the original CHIME stacking result. In addition, to test whether HyFoReS would allow us to relax the delay filter and preserve more LSS information in the data, we have experimented with three different delay cutoffs and found that the highest signal-to-noise ratio is reached at $12.22 \sigma$ using a delay cutoff that is 20 ns lower than the one in the original analysis, achieving a $5\%$ improvement.
    
Although the overall benefits from HyFoReS seem to be modest in this case, there is still significance in these results. First, we have shown that the algorithm improves detection significance in real experimental data, establishing our algorithm as a proof-of-concept foreground mitigation method. This result supplements previous studies where the formalism was used to correct antenna gain errors in simulations \citep{haochen_first_paper}, indicating the method's potential at mitigating a wide range of telescope systematics. Furthermore, we have reached a $10\sigma$ high detection significance using a delay cutoff that is 40 ns lower than the one from the original study. This, in principle, could allow the data to retain information on the third peak of BAO in Fourier space. We have thus demonstrated that, by mitigating residual foregrounds, there is hope for recovering large scale information from 21 cm data to constrain BAO, thus bringing us one step closer to CHIME's ultimate science goal.  

Despite being inferior at improving signal-to-noise on the current data compared with the outlier mask, our algorithm still has some important advantages. For example, we have found in Section~\ref{subsec:signal_loss} that signal loss due to the outlier mask increases with lower delay cutoff, but signal loss from the subtraction algorithm stays constant. This is because the algorithm by design only removes power from data that is correlated with the foregrounds. Eventually, when the delay cutoff is low enough to fully capture BAO, the control over signal loss will be critical. There are also means to improve the subtraction algorithm. For example, we can expand the correlation size or change the shape of the correlation region to account for foreground leakage from the beam's side lobes. Masking residual RFIs, excessive noise, and some of the brightest point sources will also improve the foreground residual subtraction result.

Lastly, although we have tailored the algorithm specifically for the CHIME stacking analysis, the HyFoReS formalism is general and can be applied to other 21 cm cosmology experiments as well. The algorithm works with any linear foreground filter and can in principle correct any parametrizable systematic effects on the data. In the field of 21 cm cosmology where instrument systematics are major roadblocks to its science goals, such an algorithm may prove to be useful for the community.

\begin{acknowledgements}
We thank the Dominion Radio Astrophysical Observatory, operated by the National Research Council Canada, for gracious hospitality and expertise. The DRAO is situated on the traditional, ancestral, and unceded territory of the Syilx Okanagan people. We are fortunate to live and work on these lands.

CHIME is funded by grants from the Canada Foundation for Innovation (CFI) 2012 Leading Edge Fund (Project 31170), the CFI 2015 Innovation Fund (Project 33213), and by contributions from the provinces of British Columbia, Québec, and Ontario. Long-term data storage and computational support for analysis is provided by WestGrid, SciNet and the Digital Research Alliance of Canada, and we thank their staff for flexibility and technical expertise that has been essential to this work.

Additional support was provided by the University of British Columbia, McGill University, and the University of Toronto. CHIME also benefits from NSERC Discovery Grants to several researchers, funding from the Canadian Institute for Advanced Research (CIFAR), from Canada Research Chairs, from the FRQNT Centre de Recherche en Astrophysique du Québec (CRAQ) and from the Dunlap Institute for Astronomy and Astrophysics at the University of Toronto, which is funded through an endowment established by the David Dunlap family. This material is partly based on work supported by the NSF through grants (2008031) (2006911) (2006548) and (2018490) and by the Perimeter Institute for Theoretical Physics, which in turn is supported by the Government of Canada through Industry Canada and by the Province of Ontario through the Ministry of Research and Innovation.

We thank the Sloan Digital Sky Survey and eBOSS collaborations for publicly releasing the galaxy and quasar catalogs and supporting mock catalogs used in this work.  Funding for the Sloan Digital Sky  Survey IV  has been provided by the  Alfred P. Sloan Foundation, the U.S. Department of Energy Office of  Science, and the Participating  Institutions. SDSS-IV acknowledges support and resources from the Center for High Performance Computing  at the  University of Utah. The SDSS website is www.sdss.org.

Kiyoshi Masui holds and acknowledges the support of the Adam J. Burgasser Chair in Astrophysics. Matt Dobbs is supported by a CRC Chair.
\end{acknowledgements}

\appendix
\section{REVIEW ON DECONVOLUTION IN CHIME STACKING ANALYSIS}
\label{app:decon}
In this section, we provide a summary of the CHIME stacking pipeline up to deconvolution and examine why the deconvolution procedure is unsuccessful at removing the primary beam response from the data.

The input HI data for the CHIME stacking analysis are sidereal stacks $V^q_{xy}(\nu, \phi)$, which are visibilities averaged over 102 nights of observation and stacked over redundant baselines (see full details in \citep{chime_stacking}). The index $q$ represents polarization. Each CHIME antenna feed measures two polarizations $X$ and $Y$, whose directions are orthogonal and parallel to the cylinders, respectively. Only co-polar visibilities are used in the stacking analysis, so $q \in \{XX, YY\}$. The indices $xy$ represent baselines, with $x$ and $y$ being the baseline separations orthogonal and parallel to the cylinders, respectively, or equivalently, the baseline separations along the east-west and north-south directions, respectively. The visibilites of each polarization $q$ and baseline $xy$ are functions of frequency $\nu$ and local Earth rotation angle $\phi$, which is the right ascension (RA) of the local meridian in Celestial Intermediate Reference System (CIRS).


Specific to CHIME telescope's data, it is more convenient to beamform the visibilities along the north-south direction first and leave fringing associated with east-west components of the baselines, producing hybrid beamformed visibilities. These are a function of the declination $\theta$ of the formed beam, and are related to the sky via
\begin{equation} \label{eq:beam_conv}
    \mathcal{V}^q_x(\nu, \phi, \theta) = \int \mathcal{B}^q_x(\nu,\theta,\phi - \phi^\prime) \mathcal{T}^q_x (\nu, \theta, \phi^\prime) \cos{\theta} \hbox{ } d\phi^\prime,
\end{equation}
where 
\begin{equation}
    \mathcal{T}^q_x (\nu, \theta, \phi^\prime) = \int \beta^{\hat{\theta}, q}_{\text{syn}}(\nu, \theta, \theta^\prime) S(\nu, \theta^\prime, \phi^\prime) d\theta^\prime
\end{equation}
is the flux density of the sky $S$ convolved with the north-south synthesized beam $\beta^{\hat{\theta}, q}_{\text{syn}}$, and $\mathcal{B}^q_x(\nu,\theta,\phi - \phi^\prime)$ is the beam transfer function, which is CHIME's primary beam response multiplied with a complex geometric phase factor. In the CHIME stacking analysis, the beam transfer function was constructed from an estimate of the primary beam, obtained by deconvolving a point-source-only sky model from visibilities measured by the inter-cylinder baselines (details in \citep{chime_stacking}).


The CHIME stacking analysis attempted to deconvolve the beam transfer function from the hydrid beamformed visibilities in Fourier space. Performing a Fourier transform along the $\phi$ axis of the hybrid beamformed visibilities, we obtain
\begin{equation} \label{eq:m_mode_trans}
    \tilde{\mathcal{V}}^q_x(\nu, \theta, m) = \int \mathcal{V}^q_x(\nu, \theta, \phi) e^{-im\phi} d \phi,
\end{equation}
where $m$ in Eq.~(\ref{eq:m_mode_trans}) is the Fourier dual to the local earth rotation angle $\phi$. We will refer to the operation defined in Eq.~(\ref{eq:m_mode_trans}) as the m-mode transform. The beam transfer function is also m-transformed and deconvolved from hybrid beamformed visibilities using a Tikhonov regulation scheme, 
\begin{equation} \label{eq:deconv}
    \tilde{M}^q(\nu, \theta, m) = \frac{\sum_x \mathcal{W}^q_x(\nu) \tilde{\mathcal{B}}^{q*}_x(\nu, \theta, m) \tilde{\mathcal{V}}^q_x(\nu, \theta, m)}{\eta + \sum_x \mathcal{W}^q_x(\nu) |\tilde{\mathcal{B}}^q_x(\nu, \theta, m)|^2},
\end{equation}
where $\mathcal{W}^q_x(\nu)$ is a relative weight given to each east-west baseline (more details below), and $\eta$ is a regularization parameter which takes the value of the assumed inverse signal-to-noise. Eq.~(\ref{eq:deconv}) performs a weighted average of the measurements made by the east-west baselines and deconvolves the primary beam by dividing by the weighted average of the m-mode transform of the beam transfer function, while the regularization parameter $\eta$ is there to ensure that data are divided by the beam model only when data is signal dominated. This prevents noise from being amplified by dividing the beam.

The relative weight $\mathcal{W}^q_x(\nu)$ is proportional to the inverse variance of the noise in the $m$-transformed hybrid visibility for inter-cylinder baselines (baselines with $x>0$) but is zero for intra-cylinder baselines (baselines with $x=0$). Measurements made by the intra-cylinder baselines occupy low $m$-modes of the data, which are sensitive to diffuse Galactic emissions as well as slowly-varying noise crosstalks. Therefore, Eq.~(\ref{eq:deconv}) also serves as an $m$-mode filter in the stacking pipeline to mitigate foregrounds and reduce crosstalk contamination. 

However, it is difficult to achieve complete deconvolution through Eq.~(\ref{eq:deconv}). The regularization parameter $\eta$ is chosen uniformly to be $10^{-4}$ to maximize point-source sensitivity, but the beam response, i.e., the second term on the denominator of Eq.~(\ref{eq:deconv}), is low for a large number of $m$-modes at each declination and frequency. In the sidelobe regime where $\eta \gtrsim |\tilde{\mathcal{B}}^q_x(\nu, \theta, m)|^2$, the regularization scheme of Eq.~(\ref{eq:deconv}) prevents deconvolving the east-west response. In principle, choosing lower values of $\eta$ would allow better deconvolution, but smaller regularization parameters risk amplifying the noise in the map and causing leakage in the foreground-filtered data. This subtlety with deconvolution was not directly addressed in the original CHIME stacking analysis. The primary beam response that remains in the data causes foreground residual contamination in the filtered signal, which we intend to remove using HyFoReS in this study.

\section{ESTIMATING DELAY SPECTRUM}
\label{app:dspec}
In this section, we provide details regarding the delay power spectrum and delay spectrum, and summarize how we obtain the later from the former. The delay power spectrum is defined as the variance of the delay spectrum at each delay $\tau$ and declination $\theta$. The delay power spectrum can, in principle, be estimated as 
\begin{equation}
        P(\tau,\theta) = \text{Var}_{\phi} \left\{ \tilde{m}^q(\tau,\theta,\phi) \right\},
    \end{equation}  
where $\text{Var}_{\phi} \{...\}$ represents the variance across the RA axis, and $\tilde{m}^q(\tau,\theta,\phi)$ is the delay spectrum as a function of polarization $q$, delay $\tau$, declination $\theta$, and RA $\phi$. Note that the delay power spectrum $P$ also depends on the polarization $q$, which we have suppressed in our notation for simplicity.

In the original CHIME stacking analysis, the delay power spectrum was computed to determine the declination-depend delay cutoff for the DAYENU delay foreground filter (blue curves in Fig.~\ref{fig:cutoffs}). Due to lost frequency information in the data from RFI masking, the CHIME stacking analysis used a Gibbs sampling technique to estimate the delay power spectrum. It treats the map $\bm{m}$ and delay spectrum $\bm{\tilde{m}}$ as flattened one-dimensional vectors and regards the delay power spectrum $\mbf{P}$ as a diagonal matrix with the diagonal elements being the variance of the delay spectrum $\bm{\tilde{m}}$. The Gibbs sampler then estimates the delay spectrum $\bm{\tilde{m}}$ and delay power spectrum $\mbf{P}$ simultaneously by drawing samples from a joint probability distribution $\mathcal{P}(\bm{\tilde{m}}, \mbf{P}|\bm{m})$, given the data map $\bm{m}$. The details are described in the Appendix of \citep{chime_stacking}. 

In this work, the delay power spectrum is used to qualitatively examine the results of foreground subtraction in Section~\ref{subsec:res_1}. In addition to the delay power spectrum, we also need the delay spectrum to produce estimated foregrounds and signal for HyFoReS in Section~\ref{subsec:implement}. The original CHIME stacking pipeline only saves the delay power spectrum $\mbf{P}$. To recover the delay spectrum $\bm{\tilde{m}}$, we apply a Wiener filter \citep{wiener_filter} on the map
\begin{equation} \label{eq:wiener}
    \bm{\tilde{m}} = (\mbf{P}^{-1} + \mbf{R}^\dagger \mbf{N}^{-1} \mbf{R})^{-1} \mbf{R}^\dagger \mbf{N}^{-1} \bm{m},
\end{equation}
where $\mbf{R}$ with $R_{ab} \propto e^{-2\pi j\tau_a\nu_b}$ is the discrete Fourier transform and $\mbf{N}^{-1}$ is the inverse noise variance matrix with diagonal elements corresponding to masked frequency channels being zero. Intuitively, the role of the inverse delay power spectrum $\mbf{P}^{-1}$ in Eq.~(\ref{eq:wiener}) is to give the Wiener filter statistical information on the data lost in the masked frequency channels of the frequency spectrum, resulting in significantly lower mixing of power in the delay spectrum. 

One subtlety is that Eq.~(\ref{eq:wiener}) will estimate the delay spectrum of the signal and foregrounds but exclude the noise component. In the regime of low signal-to-noise (such as at high delays where HI signal is expected to lie below the level of noise fluctuations), the Wiener filter will attempt to recover a subdominant component of the total power, therefore sensitive to errors in noise modelling. To avoid this issue, we can use Eq.~(\ref{eq:wiener}) to estimate the total power of the delay spectrum instead (including the noise component) by lowering the input noise covariance $\mbf{N}$ for the Wiener filter. This is to make sure the true noise in the data to not be down-weighted by the Wiener filter. Currently, we choose to lower the input noise covariance by a factor of $10^{-5}$, which has shown good accuracy in recovering the true delay power spectrum (including the noise power) when tested on a toy-model delay power spectrum.

\bibliography{journals,lit}

\begin{thebibliography}{28}%
\makeatletter
\providecommand \@ifxundefined [1]{%
 \@ifx{#1\undefined}
}%
\providecommand \@ifnum [1]{%
 \ifnum #1\expandafter \@firstoftwo
 \else \expandafter \@secondoftwo
 \fi
}%
\providecommand \@ifx [1]{%
 \ifx #1\expandafter \@firstoftwo
 \else \expandafter \@secondoftwo
 \fi
}%
\providecommand \natexlab [1]{#1}%
\providecommand \enquote  [1]{``#1''}%
\providecommand \bibnamefont  [1]{#1}%
\providecommand \bibfnamefont [1]{#1}%
\providecommand \citenamefont [1]{#1}%
\providecommand \href@noop [0]{\@secondoftwo}%
\providecommand \href [0]{\begingroup \@sanitize@url \@href}%
\providecommand \@href[1]{\@@startlink{#1}\@@href}%
\providecommand \@@href[1]{\endgroup#1\@@endlink}%
\providecommand \@sanitize@url [0]{\catcode `\\12\catcode `\$12\catcode `\&12\catcode `\#12\catcode `\^12\catcode `\_12\catcode `\%12\relax}%
\providecommand \@@startlink[1]{}%
\providecommand \@@endlink[0]{}%
\providecommand \url  [0]{\begingroup\@sanitize@url \@url }%
\providecommand \@url [1]{\endgroup\@href {#1}{\urlprefix }}%
\providecommand \urlprefix  [0]{URL }%
\providecommand \Eprint [0]{\href }%
\providecommand \doibase [0]{https://doi.org/}%
\providecommand \selectlanguage [0]{\@gobble}%
\providecommand \bibinfo  [0]{\@secondoftwo}%
\providecommand \bibfield  [0]{\@secondoftwo}%
\providecommand \translation [1]{[#1]}%
\providecommand \BibitemOpen [0]{}%
\providecommand \bibitemStop [0]{}%
\providecommand \bibitemNoStop [0]{.\EOS\space}%
\providecommand \EOS [0]{\spacefactor3000\relax}%
\providecommand \BibitemShut  [1]{\csname bibitem#1\endcsname}%
\let\auto@bib@innerbib\@empty
\bibitem [{\citenamefont {Chang}\ \emph {et~al.}(2008)\citenamefont {Chang}, \citenamefont {Pen}, \citenamefont {Peterson},\ and\ \citenamefont {McDonald}}]{chang_pen}%
  \BibitemOpen
  \bibfield  {author} {\bibinfo {author} {\bibfnamefont {T.-C.}\ \bibnamefont {Chang}}, \bibinfo {author} {\bibfnamefont {U.-L.}\ \bibnamefont {Pen}}, \bibinfo {author} {\bibfnamefont {J.~B.}\ \bibnamefont {Peterson}},\ and\ \bibinfo {author} {\bibfnamefont {P.}~\bibnamefont {McDonald}},\ }\bibfield  {title} {\bibinfo {title} {Baryon acoustic oscillation intensity mapping of dark energy},\ }\href {https://doi.org/10.1103/PhysRevLett.100.091303} {\bibfield  {journal} {\bibinfo  {journal} {Phys. Rev. Lett.}\ }\textbf {\bibinfo {volume} {100}},\ \bibinfo {pages} {091303} (\bibinfo {year} {2008})}\BibitemShut {NoStop}%
\bibitem [{\citenamefont {Loeb}\ and\ \citenamefont {Wyithe}(2008)}]{Loeb}%
  \BibitemOpen
  \bibfield  {author} {\bibinfo {author} {\bibfnamefont {A.}~\bibnamefont {Loeb}}\ and\ \bibinfo {author} {\bibfnamefont {J.~S.~B.}\ \bibnamefont {Wyithe}},\ }\bibfield  {title} {\bibinfo {title} {Possibility of precise measurement of the cosmological power spectrum with a dedicated survey of 21 cm emission after reionization},\ }\href {https://doi.org/10.1103/PhysRevLett.100.161301} {\bibfield  {journal} {\bibinfo  {journal} {Phys. Rev. Lett.}\ }\textbf {\bibinfo {volume} {100}},\ \bibinfo {pages} {161301} (\bibinfo {year} {2008})}\BibitemShut {NoStop}%
\bibitem [{\citenamefont {Amiri}\ \emph {et~al.}(2022)\citenamefont {Amiri}, \citenamefont {Bandura}, \citenamefont {Boskovic}, \citenamefont {Chen}, \citenamefont {Cliche}, \citenamefont {Deng}, \citenamefont {Denman}, \citenamefont {Dobbs}, \citenamefont {Fandino}, \citenamefont {Foreman}, \citenamefont {Halpern}, \citenamefont {Hanna}, \citenamefont {Hill}, \citenamefont {Hinshaw}, \citenamefont {Höfer}, \citenamefont {Kania}, \citenamefont {Klages}, \citenamefont {Landecker}, \citenamefont {MacEachern}, \citenamefont {Masui}, \citenamefont {Mena-Parra}, \citenamefont {Milutinovic}, \citenamefont {Mirhosseini}, \citenamefont {Newburgh}, \citenamefont {Nitsche}, \citenamefont {Ordog}, \citenamefont {Pen}, \citenamefont {Pinsonneault-Marotte}, \citenamefont {Polzin}, \citenamefont {Reda}, \citenamefont {Renard}, \citenamefont {Shaw}, \citenamefont {Siegel}, \citenamefont {Singh}, \citenamefont {Smegal}, \citenamefont {Tretyakov}, \citenamefont {Van~Gassen}, \citenamefont {Vanderlinde}, \citenamefont {Wang},
  \citenamefont {Wiebe}, \citenamefont {Willis},\ and\ \citenamefont {Wulf}}]{chime_overview}%
  \BibitemOpen
  \bibfield  {author} {\bibinfo {author} {\bibfnamefont {M.}~\bibnamefont {Amiri}}, \bibinfo {author} {\bibfnamefont {K.}~\bibnamefont {Bandura}}, \bibinfo {author} {\bibfnamefont {A.}~\bibnamefont {Boskovic}}, \bibinfo {author} {\bibfnamefont {T.}~\bibnamefont {Chen}}, \bibinfo {author} {\bibfnamefont {J.-F.}\ \bibnamefont {Cliche}}, \bibinfo {author} {\bibfnamefont {M.}~\bibnamefont {Deng}}, \bibinfo {author} {\bibfnamefont {N.}~\bibnamefont {Denman}}, \bibinfo {author} {\bibfnamefont {M.}~\bibnamefont {Dobbs}}, \bibinfo {author} {\bibfnamefont {M.}~\bibnamefont {Fandino}}, \bibinfo {author} {\bibfnamefont {S.}~\bibnamefont {Foreman}}, \bibinfo {author} {\bibfnamefont {M.}~\bibnamefont {Halpern}}, \bibinfo {author} {\bibfnamefont {D.}~\bibnamefont {Hanna}}, \bibinfo {author} {\bibfnamefont {A.~S.}\ \bibnamefont {Hill}}, \bibinfo {author} {\bibfnamefont {G.}~\bibnamefont {Hinshaw}}, \bibinfo {author} {\bibfnamefont {C.}~\bibnamefont {Höfer}}, \bibinfo {author} {\bibfnamefont {J.}~\bibnamefont {Kania}},
  \bibinfo {author} {\bibfnamefont {P.}~\bibnamefont {Klages}}, \bibinfo {author} {\bibfnamefont {T.~L.}\ \bibnamefont {Landecker}}, \bibinfo {author} {\bibfnamefont {J.}~\bibnamefont {MacEachern}}, \bibinfo {author} {\bibfnamefont {K.}~\bibnamefont {Masui}}, \bibinfo {author} {\bibfnamefont {J.}~\bibnamefont {Mena-Parra}}, \bibinfo {author} {\bibfnamefont {N.}~\bibnamefont {Milutinovic}}, \bibinfo {author} {\bibfnamefont {A.}~\bibnamefont {Mirhosseini}}, \bibinfo {author} {\bibfnamefont {L.}~\bibnamefont {Newburgh}}, \bibinfo {author} {\bibfnamefont {R.}~\bibnamefont {Nitsche}}, \bibinfo {author} {\bibfnamefont {A.}~\bibnamefont {Ordog}}, \bibinfo {author} {\bibfnamefont {U.-L.}\ \bibnamefont {Pen}}, \bibinfo {author} {\bibfnamefont {T.}~\bibnamefont {Pinsonneault-Marotte}}, \bibinfo {author} {\bibfnamefont {A.}~\bibnamefont {Polzin}}, \bibinfo {author} {\bibfnamefont {A.}~\bibnamefont {Reda}}, \bibinfo {author} {\bibfnamefont {A.}~\bibnamefont {Renard}}, \bibinfo {author} {\bibfnamefont {J.~R.}\
  \bibnamefont {Shaw}}, \bibinfo {author} {\bibfnamefont {S.~R.}\ \bibnamefont {Siegel}}, \bibinfo {author} {\bibfnamefont {S.}~\bibnamefont {Singh}}, \bibinfo {author} {\bibfnamefont {R.}~\bibnamefont {Smegal}}, \bibinfo {author} {\bibfnamefont {I.}~\bibnamefont {Tretyakov}}, \bibinfo {author} {\bibfnamefont {K.}~\bibnamefont {Van~Gassen}}, \bibinfo {author} {\bibfnamefont {K.}~\bibnamefont {Vanderlinde}}, \bibinfo {author} {\bibfnamefont {H.}~\bibnamefont {Wang}}, \bibinfo {author} {\bibfnamefont {D.~V.}\ \bibnamefont {Wiebe}}, \bibinfo {author} {\bibfnamefont {J.~S.}\ \bibnamefont {Willis}},\ and\ \bibinfo {author} {\bibfnamefont {D.}~\bibnamefont {Wulf}},\ }\bibfield  {title} {\bibinfo {title} {An overview of chime, the canadian hydrogen intensity mapping experiment},\ }\href {https://doi.org/10.3847/1538-4365/ac6fd9} {\bibfield  {journal} {\bibinfo  {journal} {The Astrophysical Journal Supplement Series}\ }\textbf {\bibinfo {volume} {261}},\ \bibinfo {pages} {29} (\bibinfo {year} {2022})}\BibitemShut
  {NoStop}%
\bibitem [{\citenamefont {Santos}\ \emph {et~al.}(2017)\citenamefont {Santos}, \citenamefont {Cluver}, \citenamefont {Hilton}, \citenamefont {Jarvis}, \citenamefont {Jozsa}, \citenamefont {Leeuw}, \citenamefont {Smirnov}, \citenamefont {Taylor}, \citenamefont {Abdalla}, \citenamefont {Afonso}, \citenamefont {Alonso}, \citenamefont {Bacon}, \citenamefont {Bassett}, \citenamefont {Bernardi}, \citenamefont {Bull}, \citenamefont {Camera}, \citenamefont {Chiang}, \citenamefont {Colafrancesco}, \citenamefont {Ferreira}, \citenamefont {Fonseca}, \citenamefont {van~der Heyden}, \citenamefont {Heywood}, \citenamefont {Knowles}, \citenamefont {Lochner}, \citenamefont {Ma}, \citenamefont {Maartens}, \citenamefont {Makhathini}, \citenamefont {Moodley}, \citenamefont {Pourtsidou}, \citenamefont {Prescott}, \citenamefont {Sievers}, \citenamefont {Spekkens}, \citenamefont {Vaccari}, \citenamefont {Weltman}, \citenamefont {Whittam}, \citenamefont {Witzemann}, \citenamefont {Wolz},\ and\ \citenamefont {Zwart}}]{meerkat}%
  \BibitemOpen
  \bibfield  {author} {\bibinfo {author} {\bibfnamefont {M.~G.}\ \bibnamefont {Santos}}, \bibinfo {author} {\bibfnamefont {M.}~\bibnamefont {Cluver}}, \bibinfo {author} {\bibfnamefont {M.}~\bibnamefont {Hilton}}, \bibinfo {author} {\bibfnamefont {M.}~\bibnamefont {Jarvis}}, \bibinfo {author} {\bibfnamefont {G.~I.~G.}\ \bibnamefont {Jozsa}}, \bibinfo {author} {\bibfnamefont {L.}~\bibnamefont {Leeuw}}, \bibinfo {author} {\bibfnamefont {O.}~\bibnamefont {Smirnov}}, \bibinfo {author} {\bibfnamefont {R.}~\bibnamefont {Taylor}}, \bibinfo {author} {\bibfnamefont {F.}~\bibnamefont {Abdalla}}, \bibinfo {author} {\bibfnamefont {J.}~\bibnamefont {Afonso}}, \bibinfo {author} {\bibfnamefont {D.}~\bibnamefont {Alonso}}, \bibinfo {author} {\bibfnamefont {D.}~\bibnamefont {Bacon}}, \bibinfo {author} {\bibfnamefont {B.~A.}\ \bibnamefont {Bassett}}, \bibinfo {author} {\bibfnamefont {G.}~\bibnamefont {Bernardi}}, \bibinfo {author} {\bibfnamefont {P.}~\bibnamefont {Bull}}, \bibinfo {author} {\bibfnamefont {S.}~\bibnamefont
  {Camera}}, \bibinfo {author} {\bibfnamefont {H.~C.}\ \bibnamefont {Chiang}}, \bibinfo {author} {\bibfnamefont {S.}~\bibnamefont {Colafrancesco}}, \bibinfo {author} {\bibfnamefont {P.~G.}\ \bibnamefont {Ferreira}}, \bibinfo {author} {\bibfnamefont {J.}~\bibnamefont {Fonseca}}, \bibinfo {author} {\bibfnamefont {K.}~\bibnamefont {van~der Heyden}}, \bibinfo {author} {\bibfnamefont {I.}~\bibnamefont {Heywood}}, \bibinfo {author} {\bibfnamefont {K.}~\bibnamefont {Knowles}}, \bibinfo {author} {\bibfnamefont {M.}~\bibnamefont {Lochner}}, \bibinfo {author} {\bibfnamefont {Y.-Z.}\ \bibnamefont {Ma}}, \bibinfo {author} {\bibfnamefont {R.}~\bibnamefont {Maartens}}, \bibinfo {author} {\bibfnamefont {S.}~\bibnamefont {Makhathini}}, \bibinfo {author} {\bibfnamefont {K.}~\bibnamefont {Moodley}}, \bibinfo {author} {\bibfnamefont {A.}~\bibnamefont {Pourtsidou}}, \bibinfo {author} {\bibfnamefont {M.}~\bibnamefont {Prescott}}, \bibinfo {author} {\bibfnamefont {J.}~\bibnamefont {Sievers}}, \bibinfo {author} {\bibfnamefont
  {K.}~\bibnamefont {Spekkens}}, \bibinfo {author} {\bibfnamefont {M.}~\bibnamefont {Vaccari}}, \bibinfo {author} {\bibfnamefont {A.}~\bibnamefont {Weltman}}, \bibinfo {author} {\bibfnamefont {I.}~\bibnamefont {Whittam}}, \bibinfo {author} {\bibfnamefont {A.}~\bibnamefont {Witzemann}}, \bibinfo {author} {\bibfnamefont {L.}~\bibnamefont {Wolz}},\ and\ \bibinfo {author} {\bibfnamefont {J.~T.~L.}\ \bibnamefont {Zwart}},\ }\href@noop {} {\bibinfo {title} {Meerklass: Meerkat large area synoptic survey}} (\bibinfo {year} {2017}),\ \Eprint {https://arxiv.org/abs/1709.06099} {arXiv:1709.06099 [astro-ph.CO]} \BibitemShut {NoStop}%
\bibitem [{\citenamefont {Chakraborty}\ \emph {et~al.}(2021)\citenamefont {Chakraborty}, \citenamefont {Datta}, \citenamefont {Roy}, \citenamefont {Bharadwaj}, \citenamefont {Choudhury}, \citenamefont {Datta}, \citenamefont {Pal}, \citenamefont {Choudhury}, \citenamefont {Choudhuri}, \citenamefont {Dutta},\ and\ \citenamefont {Sarkar}}]{ugmrt}%
  \BibitemOpen
  \bibfield  {author} {\bibinfo {author} {\bibfnamefont {A.}~\bibnamefont {Chakraborty}}, \bibinfo {author} {\bibfnamefont {A.}~\bibnamefont {Datta}}, \bibinfo {author} {\bibfnamefont {N.}~\bibnamefont {Roy}}, \bibinfo {author} {\bibfnamefont {S.}~\bibnamefont {Bharadwaj}}, \bibinfo {author} {\bibfnamefont {T.~R.}\ \bibnamefont {Choudhury}}, \bibinfo {author} {\bibfnamefont {K.~K.}\ \bibnamefont {Datta}}, \bibinfo {author} {\bibfnamefont {S.}~\bibnamefont {Pal}}, \bibinfo {author} {\bibfnamefont {M.}~\bibnamefont {Choudhury}}, \bibinfo {author} {\bibfnamefont {S.}~\bibnamefont {Choudhuri}}, \bibinfo {author} {\bibfnamefont {P.}~\bibnamefont {Dutta}},\ and\ \bibinfo {author} {\bibfnamefont {D.}~\bibnamefont {Sarkar}},\ }\bibfield  {title} {\bibinfo {title} {First multi-redshift limits on post–epoch of reionization 21 cm signal from z = 1.96–3.58 using ugmrt},\ }\href {https://doi.org/10.3847/2041-8213/abd17a} {\bibfield  {journal} {\bibinfo  {journal} {The Astrophysical Journal Letters}\ }\textbf {\bibinfo
  {volume} {907}},\ \bibinfo {pages} {L7} (\bibinfo {year} {2021})}\BibitemShut {NoStop}%
\bibitem [{\citenamefont {DeBoer}\ \emph {et~al.}(2017)\citenamefont {DeBoer}, \citenamefont {Parsons}, \citenamefont {Aguirre} \emph {et~al.}}]{HERA_2017a}%
  \BibitemOpen
  \bibfield  {author} {\bibinfo {author} {\bibfnamefont {D.~R.}\ \bibnamefont {DeBoer}}, \bibinfo {author} {\bibfnamefont {A.~R.}\ \bibnamefont {Parsons}}, \bibinfo {author} {\bibfnamefont {J.~E.}\ \bibnamefont {Aguirre}}, \emph {et~al.},\ }\bibfield  {title} {\bibinfo {title} {Hydrogen epoch of reionization array ({HERA})},\ }\href {https://doi.org/10.1088/1538-3873/129/974/045001} {\bibfield  {journal} {\bibinfo  {journal} {The Publications of the Astronomical Society of the Pacific}\ }\textbf {\bibinfo {volume} {129}},\ \bibinfo {pages} {045001} (\bibinfo {year} {2017})}\BibitemShut {NoStop}%
\bibitem [{\citenamefont {Patil}\ \emph {et~al.}(2017)\citenamefont {Patil}, \citenamefont {Yatawatta}, \citenamefont {Koopmans} \emph {et~al.}}]{LOFAR_2017}%
  \BibitemOpen
  \bibfield  {author} {\bibinfo {author} {\bibfnamefont {A.~H.}\ \bibnamefont {Patil}}, \bibinfo {author} {\bibfnamefont {S.}~\bibnamefont {Yatawatta}}, \bibinfo {author} {\bibfnamefont {L.~V.~E.}\ \bibnamefont {Koopmans}}, \emph {et~al.},\ }\bibfield  {title} {\bibinfo {title} {Upper limits on the 21 cm epoch of reionization power spectrum from one night with {LOFAR}},\ }\href {https://doi.org/10.3847/1538-4357/aa63e7} {\bibfield  {journal} {\bibinfo  {journal} {The Astrophysical Journal}\ }\textbf {\bibinfo {volume} {838}},\ \bibinfo {pages} {65} (\bibinfo {year} {2017})}\BibitemShut {NoStop}%
\bibitem [{\citenamefont {Barry}\ \emph {et~al.}(2019)\citenamefont {Barry}, \citenamefont {Wilensky}, \citenamefont {Trott} \emph {et~al.}}]{MWA_2019}%
  \BibitemOpen
  \bibfield  {author} {\bibinfo {author} {\bibfnamefont {N.}~\bibnamefont {Barry}}, \bibinfo {author} {\bibfnamefont {M.}~\bibnamefont {Wilensky}}, \bibinfo {author} {\bibfnamefont {C.~M.}\ \bibnamefont {Trott}}, \emph {et~al.},\ }\bibfield  {title} {\bibinfo {title} {Improving the epoch of reionization power spectrum results from murchison widefield array season 1 observations},\ }\href {https://doi.org/10.3847/1538-4357/ab40a8} {\bibfield  {journal} {\bibinfo  {journal} {The Astrophysical Journal}\ }\textbf {\bibinfo {volume} {884}},\ \bibinfo {pages} {1} (\bibinfo {year} {2019})}\BibitemShut {NoStop}%
\bibitem [{\citenamefont {Collaboration}\ \emph {et~al.}(2023)\citenamefont {Collaboration}, \citenamefont {Amiri}, \citenamefont {Bandura}, \citenamefont {Chakraborty}, \citenamefont {Dobbs}, \citenamefont {Fandino}, \citenamefont {Foreman}, \citenamefont {Gan}, \citenamefont {Halpern}, \citenamefont {Hill}, \citenamefont {Hinshaw}, \citenamefont {Höfer}, \citenamefont {Landecker}, \citenamefont {Li}, \citenamefont {MacEachern}, \citenamefont {Masui}, \citenamefont {Mena-Parra}, \citenamefont {Milutinovic}, \citenamefont {Mirhosseini}, \citenamefont {Newburgh}, \citenamefont {Ordog}, \citenamefont {Paul}, \citenamefont {Pen}, \citenamefont {Pinsonneault-Marotte}, \citenamefont {Reda}, \citenamefont {Shaw}, \citenamefont {Siegel}, \citenamefont {Vanderlinde}, \citenamefont {Wang}, \citenamefont {Wiebe},\ and\ \citenamefont {Wulf}}]{chime_lyman}%
  \BibitemOpen
  \bibfield  {author} {\bibinfo {author} {\bibfnamefont {C.}~\bibnamefont {Collaboration}}, \bibinfo {author} {\bibfnamefont {M.}~\bibnamefont {Amiri}}, \bibinfo {author} {\bibfnamefont {K.}~\bibnamefont {Bandura}}, \bibinfo {author} {\bibfnamefont {A.}~\bibnamefont {Chakraborty}}, \bibinfo {author} {\bibfnamefont {M.}~\bibnamefont {Dobbs}}, \bibinfo {author} {\bibfnamefont {M.}~\bibnamefont {Fandino}}, \bibinfo {author} {\bibfnamefont {S.}~\bibnamefont {Foreman}}, \bibinfo {author} {\bibfnamefont {H.}~\bibnamefont {Gan}}, \bibinfo {author} {\bibfnamefont {M.}~\bibnamefont {Halpern}}, \bibinfo {author} {\bibfnamefont {A.~S.}\ \bibnamefont {Hill}}, \bibinfo {author} {\bibfnamefont {G.}~\bibnamefont {Hinshaw}}, \bibinfo {author} {\bibfnamefont {C.}~\bibnamefont {Höfer}}, \bibinfo {author} {\bibfnamefont {T.~L.}\ \bibnamefont {Landecker}}, \bibinfo {author} {\bibfnamefont {Z.}~\bibnamefont {Li}}, \bibinfo {author} {\bibfnamefont {J.}~\bibnamefont {MacEachern}}, \bibinfo {author} {\bibfnamefont {K.}~\bibnamefont
  {Masui}}, \bibinfo {author} {\bibfnamefont {J.}~\bibnamefont {Mena-Parra}}, \bibinfo {author} {\bibfnamefont {N.}~\bibnamefont {Milutinovic}}, \bibinfo {author} {\bibfnamefont {A.}~\bibnamefont {Mirhosseini}}, \bibinfo {author} {\bibfnamefont {L.}~\bibnamefont {Newburgh}}, \bibinfo {author} {\bibfnamefont {A.}~\bibnamefont {Ordog}}, \bibinfo {author} {\bibfnamefont {S.}~\bibnamefont {Paul}}, \bibinfo {author} {\bibfnamefont {U.-L.}\ \bibnamefont {Pen}}, \bibinfo {author} {\bibfnamefont {T.}~\bibnamefont {Pinsonneault-Marotte}}, \bibinfo {author} {\bibfnamefont {A.}~\bibnamefont {Reda}}, \bibinfo {author} {\bibfnamefont {J.~R.}\ \bibnamefont {Shaw}}, \bibinfo {author} {\bibfnamefont {S.~R.}\ \bibnamefont {Siegel}}, \bibinfo {author} {\bibfnamefont {K.}~\bibnamefont {Vanderlinde}}, \bibinfo {author} {\bibfnamefont {H.}~\bibnamefont {Wang}}, \bibinfo {author} {\bibfnamefont {D.~V.}\ \bibnamefont {Wiebe}},\ and\ \bibinfo {author} {\bibfnamefont {D.}~\bibnamefont {Wulf}},\ }\href@noop {} {\bibinfo {title} {A
  detection of cosmological 21 cm emission from chime in cross-correlation with eboss measurements of the lyman-$\alpha$ forest}} (\bibinfo {year} {2023}),\ \Eprint {https://arxiv.org/abs/2309.04404} {arXiv:2309.04404 [astro-ph.CO]} \BibitemShut {NoStop}%
\bibitem [{\citenamefont {Amiri}\ \emph {et~al.}(2023)\citenamefont {Amiri}, \citenamefont {Bandura}, \citenamefont {Chen}, \citenamefont {Deng}, \citenamefont {Dobbs}, \citenamefont {Fandino}, \citenamefont {Foreman}, \citenamefont {Halpern}, \citenamefont {Hill}, \citenamefont {Hinshaw}, \citenamefont {Höfer}, \citenamefont {Kania}, \citenamefont {Landecker}, \citenamefont {MacEachern}, \citenamefont {Masui}, \citenamefont {Mena-Parra}, \citenamefont {Milutinovic}, \citenamefont {Mirhosseini}, \citenamefont {Newburgh}, \citenamefont {Ordog}, \citenamefont {Pen}, \citenamefont {Pinsonneault-Marotte}, \citenamefont {Polzin}, \citenamefont {Reda}, \citenamefont {Renard}, \citenamefont {Shaw}, \citenamefont {Siegel}, \citenamefont {Singh}, \citenamefont {Vanderlinde}, \citenamefont {Wang}, \citenamefont {Wiebe},\ and\ \citenamefont {Wulf}}]{chime_stacking}%
  \BibitemOpen
  \bibfield  {author} {\bibinfo {author} {\bibfnamefont {M.}~\bibnamefont {Amiri}}, \bibinfo {author} {\bibfnamefont {K.}~\bibnamefont {Bandura}}, \bibinfo {author} {\bibfnamefont {T.}~\bibnamefont {Chen}}, \bibinfo {author} {\bibfnamefont {M.}~\bibnamefont {Deng}}, \bibinfo {author} {\bibfnamefont {M.}~\bibnamefont {Dobbs}}, \bibinfo {author} {\bibfnamefont {M.}~\bibnamefont {Fandino}}, \bibinfo {author} {\bibfnamefont {S.}~\bibnamefont {Foreman}}, \bibinfo {author} {\bibfnamefont {M.}~\bibnamefont {Halpern}}, \bibinfo {author} {\bibfnamefont {A.~S.}\ \bibnamefont {Hill}}, \bibinfo {author} {\bibfnamefont {G.}~\bibnamefont {Hinshaw}}, \bibinfo {author} {\bibfnamefont {C.}~\bibnamefont {Höfer}}, \bibinfo {author} {\bibfnamefont {J.}~\bibnamefont {Kania}}, \bibinfo {author} {\bibfnamefont {T.~L.}\ \bibnamefont {Landecker}}, \bibinfo {author} {\bibfnamefont {J.}~\bibnamefont {MacEachern}}, \bibinfo {author} {\bibfnamefont {K.}~\bibnamefont {Masui}}, \bibinfo {author} {\bibfnamefont {J.}~\bibnamefont
  {Mena-Parra}}, \bibinfo {author} {\bibfnamefont {N.}~\bibnamefont {Milutinovic}}, \bibinfo {author} {\bibfnamefont {A.}~\bibnamefont {Mirhosseini}}, \bibinfo {author} {\bibfnamefont {L.}~\bibnamefont {Newburgh}}, \bibinfo {author} {\bibfnamefont {A.}~\bibnamefont {Ordog}}, \bibinfo {author} {\bibfnamefont {U.-L.}\ \bibnamefont {Pen}}, \bibinfo {author} {\bibfnamefont {T.}~\bibnamefont {Pinsonneault-Marotte}}, \bibinfo {author} {\bibfnamefont {A.}~\bibnamefont {Polzin}}, \bibinfo {author} {\bibfnamefont {A.}~\bibnamefont {Reda}}, \bibinfo {author} {\bibfnamefont {A.}~\bibnamefont {Renard}}, \bibinfo {author} {\bibfnamefont {J.~R.}\ \bibnamefont {Shaw}}, \bibinfo {author} {\bibfnamefont {S.~R.}\ \bibnamefont {Siegel}}, \bibinfo {author} {\bibfnamefont {S.}~\bibnamefont {Singh}}, \bibinfo {author} {\bibfnamefont {K.}~\bibnamefont {Vanderlinde}}, \bibinfo {author} {\bibfnamefont {H.}~\bibnamefont {Wang}}, \bibinfo {author} {\bibfnamefont {D.~V.}\ \bibnamefont {Wiebe}},\ and\ \bibinfo {author} {\bibfnamefont
  {D.}~\bibnamefont {Wulf}},\ }\bibfield  {title} {\bibinfo {title} {Detection of cosmological 21 cm emission with the canadian hydrogen intensity mapping experiment},\ }\href {https://doi.org/10.3847/1538-4357/acb13f} {\bibfield  {journal} {\bibinfo  {journal} {The Astrophysical Journal}\ }\textbf {\bibinfo {volume} {947}},\ \bibinfo {pages} {16} (\bibinfo {year} {2023})}\BibitemShut {NoStop}%
\bibitem [{\citenamefont {Wolz}\ \emph {et~al.}(2021)\citenamefont {Wolz}, \citenamefont {Pourtsidou}, \citenamefont {Masui}, \citenamefont {Chang}, \citenamefont {Bautista}, \citenamefont {Müller}, \citenamefont {Avila}, \citenamefont {Bacon}, \citenamefont {Percival}, \citenamefont {Cunnington}, \citenamefont {Anderson}, \citenamefont {Chen}, \citenamefont {Kneib}, \citenamefont {Li}, \citenamefont {Liao}, \citenamefont {Pen}, \citenamefont {Peterson}, \citenamefont {Rossi}, \citenamefont {Schneider}, \citenamefont {Yadav},\ and\ \citenamefont {Zhao}}]{GBT_cross}%
  \BibitemOpen
  \bibfield  {author} {\bibinfo {author} {\bibfnamefont {L.}~\bibnamefont {Wolz}}, \bibinfo {author} {\bibfnamefont {A.}~\bibnamefont {Pourtsidou}}, \bibinfo {author} {\bibfnamefont {K.~W.}\ \bibnamefont {Masui}}, \bibinfo {author} {\bibfnamefont {T.-C.}\ \bibnamefont {Chang}}, \bibinfo {author} {\bibfnamefont {J.~E.}\ \bibnamefont {Bautista}}, \bibinfo {author} {\bibfnamefont {E.-M.}\ \bibnamefont {Müller}}, \bibinfo {author} {\bibfnamefont {S.}~\bibnamefont {Avila}}, \bibinfo {author} {\bibfnamefont {D.}~\bibnamefont {Bacon}}, \bibinfo {author} {\bibfnamefont {W.~J.}\ \bibnamefont {Percival}}, \bibinfo {author} {\bibfnamefont {S.}~\bibnamefont {Cunnington}}, \bibinfo {author} {\bibfnamefont {C.}~\bibnamefont {Anderson}}, \bibinfo {author} {\bibfnamefont {X.}~\bibnamefont {Chen}}, \bibinfo {author} {\bibfnamefont {J.-P.}\ \bibnamefont {Kneib}}, \bibinfo {author} {\bibfnamefont {Y.-C.}\ \bibnamefont {Li}}, \bibinfo {author} {\bibfnamefont {Y.-W.}\ \bibnamefont {Liao}}, \bibinfo {author} {\bibfnamefont
  {U.-L.}\ \bibnamefont {Pen}}, \bibinfo {author} {\bibfnamefont {J.~B.}\ \bibnamefont {Peterson}}, \bibinfo {author} {\bibfnamefont {G.}~\bibnamefont {Rossi}}, \bibinfo {author} {\bibfnamefont {D.~P.}\ \bibnamefont {Schneider}}, \bibinfo {author} {\bibfnamefont {J.}~\bibnamefont {Yadav}},\ and\ \bibinfo {author} {\bibfnamefont {G.-B.}\ \bibnamefont {Zhao}},\ }\bibfield  {title} {\bibinfo {title} {{HI constraints from the cross-correlation of eBOSS galaxies and Green Bank Telescope intensity maps}},\ }\href {https://doi.org/10.1093/mnras/stab3621} {\bibfield  {journal} {\bibinfo  {journal} {Monthly Notices of the Royal Astronomical Society}\ }\textbf {\bibinfo {volume} {510}},\ \bibinfo {pages} {3495} (\bibinfo {year} {2021})},\ \Eprint {https://arxiv.org/abs/https://academic.oup.com/mnras/article-pdf/510/3/3495/42147486/stab3621.pdf} {https://academic.oup.com/mnras/article-pdf/510/3/3495/42147486/stab3621.pdf} \BibitemShut {NoStop}%
\bibitem [{\citenamefont {Cunnington}\ \emph {et~al.}(2022)\citenamefont {Cunnington}, \citenamefont {Li}, \citenamefont {Santos}, \citenamefont {Wang}, \citenamefont {Carucci}, \citenamefont {Irfan}, \citenamefont {Pourtsidou}, \citenamefont {Spinelli}, \citenamefont {Wolz}, \citenamefont {Soares}, \citenamefont {Blake}, \citenamefont {Bull}, \citenamefont {Engelbrecht}, \citenamefont {Fonseca}, \citenamefont {Grainge},\ and\ \citenamefont {Ma}}]{meerkat_cross}%
  \BibitemOpen
  \bibfield  {author} {\bibinfo {author} {\bibfnamefont {S.}~\bibnamefont {Cunnington}}, \bibinfo {author} {\bibfnamefont {Y.}~\bibnamefont {Li}}, \bibinfo {author} {\bibfnamefont {M.~G.}\ \bibnamefont {Santos}}, \bibinfo {author} {\bibfnamefont {J.}~\bibnamefont {Wang}}, \bibinfo {author} {\bibfnamefont {I.~P.}\ \bibnamefont {Carucci}}, \bibinfo {author} {\bibfnamefont {M.~O.}\ \bibnamefont {Irfan}}, \bibinfo {author} {\bibfnamefont {A.}~\bibnamefont {Pourtsidou}}, \bibinfo {author} {\bibfnamefont {M.}~\bibnamefont {Spinelli}}, \bibinfo {author} {\bibfnamefont {L.}~\bibnamefont {Wolz}}, \bibinfo {author} {\bibfnamefont {P.~S.}\ \bibnamefont {Soares}}, \bibinfo {author} {\bibfnamefont {C.}~\bibnamefont {Blake}}, \bibinfo {author} {\bibfnamefont {P.}~\bibnamefont {Bull}}, \bibinfo {author} {\bibfnamefont {B.}~\bibnamefont {Engelbrecht}}, \bibinfo {author} {\bibfnamefont {J.}~\bibnamefont {Fonseca}}, \bibinfo {author} {\bibfnamefont {K.}~\bibnamefont {Grainge}},\ and\ \bibinfo {author} {\bibfnamefont {Y.-Z.}\
  \bibnamefont {Ma}},\ }\bibfield  {title} {\bibinfo {title} {Hi intensity mapping with meerkat: power spectrum detection in cross-correlation with wigglez galaxies},\ }\href {https://doi.org/10.1093/mnras/stac3060} {\bibfield  {journal} {\bibinfo  {journal} {Monthly Notices of the Royal Astronomical Society}\ }\textbf {\bibinfo {volume} {518}},\ \bibinfo {pages} {6262–6272} (\bibinfo {year} {2022})}\BibitemShut {NoStop}%
\bibitem [{\citenamefont {Li}\ \emph {et~al.}(2021)\citenamefont {Li}, \citenamefont {Staveley-Smith},\ and\ \citenamefont {Rhee}}]{Li_2021}%
  \BibitemOpen
  \bibfield  {author} {\bibinfo {author} {\bibfnamefont {L.-C.}\ \bibnamefont {Li}}, \bibinfo {author} {\bibfnamefont {L.}~\bibnamefont {Staveley-Smith}},\ and\ \bibinfo {author} {\bibfnamefont {J.}~\bibnamefont {Rhee}},\ }\bibfield  {title} {\bibinfo {title} {An hi intensity mapping survey with a phased array feed},\ }\href {https://doi.org/10.1088/1674-4527/21/2/30} {\bibfield  {journal} {\bibinfo  {journal} {Research in Astronomy and Astrophysics}\ }\textbf {\bibinfo {volume} {21}},\ \bibinfo {pages} {030} (\bibinfo {year} {2021})}\BibitemShut {NoStop}%
\bibitem [{\citenamefont {{Anderson}}\ \emph {et~al.}(2018)\citenamefont {{Anderson}}, \citenamefont {{Luciw}}, \citenamefont {{Li}}, \citenamefont {{Kuo}}, \citenamefont {{Yadav}}, \citenamefont {{Masui}}, \citenamefont {{Chang}}, \citenamefont {{Chen}}, \citenamefont {{Oppermann}}, \citenamefont {{Liao}}, \citenamefont {{Pen}}, \citenamefont {{Price}}, \citenamefont {{Staveley-Smith}}, \citenamefont {{Switzer}}, \citenamefont {{Timbie}},\ and\ \citenamefont {{Wolz}}}]{Anderson_2018}%
  \BibitemOpen
  \bibfield  {author} {\bibinfo {author} {\bibfnamefont {C.~J.}\ \bibnamefont {{Anderson}}}, \bibinfo {author} {\bibfnamefont {N.~J.}\ \bibnamefont {{Luciw}}}, \bibinfo {author} {\bibfnamefont {Y.~C.}\ \bibnamefont {{Li}}}, \bibinfo {author} {\bibfnamefont {C.~Y.}\ \bibnamefont {{Kuo}}}, \bibinfo {author} {\bibfnamefont {J.}~\bibnamefont {{Yadav}}}, \bibinfo {author} {\bibfnamefont {K.~W.}\ \bibnamefont {{Masui}}}, \bibinfo {author} {\bibfnamefont {T.~C.}\ \bibnamefont {{Chang}}}, \bibinfo {author} {\bibfnamefont {X.}~\bibnamefont {{Chen}}}, \bibinfo {author} {\bibfnamefont {N.}~\bibnamefont {{Oppermann}}}, \bibinfo {author} {\bibfnamefont {Y.~W.}\ \bibnamefont {{Liao}}}, \bibinfo {author} {\bibfnamefont {U.~L.}\ \bibnamefont {{Pen}}}, \bibinfo {author} {\bibfnamefont {D.~C.}\ \bibnamefont {{Price}}}, \bibinfo {author} {\bibfnamefont {L.}~\bibnamefont {{Staveley-Smith}}}, \bibinfo {author} {\bibfnamefont {E.~R.}\ \bibnamefont {{Switzer}}}, \bibinfo {author} {\bibfnamefont {P.~T.}\ \bibnamefont {{Timbie}}},\
  and\ \bibinfo {author} {\bibfnamefont {L.}~\bibnamefont {{Wolz}}},\ }\bibfield  {title} {\bibinfo {title} {{Low-amplitude clustering in low-redshift 21-cm intensity maps cross-correlated with 2dF galaxy densities}},\ }\href {https://doi.org/10.1093/mnras/sty346} {\bibfield  {journal} {\bibinfo  {journal} {MNRAS}\ }\textbf {\bibinfo {volume} {476}},\ \bibinfo {pages} {3382} (\bibinfo {year} {2018})},\ \Eprint {https://arxiv.org/abs/1710.00424} {arXiv:1710.00424 [astro-ph.CO]} \BibitemShut {NoStop}%
\bibitem [{\citenamefont {Masui}\ \emph {et~al.}(2013)\citenamefont {Masui}, \citenamefont {Switzer}, \citenamefont {Banavar}, \citenamefont {Bandura}, \citenamefont {Blake}, \citenamefont {Calin}, \citenamefont {Chang}, \citenamefont {Chen}, \citenamefont {Li}, \citenamefont {Liao}, \citenamefont {Natarajan}, \citenamefont {Pen}, \citenamefont {Peterson}, \citenamefont {Shaw},\ and\ \citenamefont {Voytek}}]{kiyo_survey}%
  \BibitemOpen
  \bibfield  {author} {\bibinfo {author} {\bibfnamefont {K.~W.}\ \bibnamefont {Masui}}, \bibinfo {author} {\bibfnamefont {E.~R.}\ \bibnamefont {Switzer}}, \bibinfo {author} {\bibfnamefont {N.}~\bibnamefont {Banavar}}, \bibinfo {author} {\bibfnamefont {K.}~\bibnamefont {Bandura}}, \bibinfo {author} {\bibfnamefont {C.}~\bibnamefont {Blake}}, \bibinfo {author} {\bibfnamefont {L.-M.}\ \bibnamefont {Calin}}, \bibinfo {author} {\bibfnamefont {T.-C.}\ \bibnamefont {Chang}}, \bibinfo {author} {\bibfnamefont {X.}~\bibnamefont {Chen}}, \bibinfo {author} {\bibfnamefont {Y.-C.}\ \bibnamefont {Li}}, \bibinfo {author} {\bibfnamefont {Y.-W.}\ \bibnamefont {Liao}}, \bibinfo {author} {\bibfnamefont {A.}~\bibnamefont {Natarajan}}, \bibinfo {author} {\bibfnamefont {U.-L.}\ \bibnamefont {Pen}}, \bibinfo {author} {\bibfnamefont {J.~B.}\ \bibnamefont {Peterson}}, \bibinfo {author} {\bibfnamefont {J.~R.}\ \bibnamefont {Shaw}},\ and\ \bibinfo {author} {\bibfnamefont {T.~C.}\ \bibnamefont {Voytek}},\ }\bibfield  {title} {\bibinfo
  {title} {Measurement of 21 cm brightness fluctuations at z $\sim$ 0.8 in cross-correlation},\ }\href {https://doi.org/10.1088/2041-8205/763/1/l20} {\bibfield  {journal} {\bibinfo  {journal} {The Astrophysical Journal}\ }\textbf {\bibinfo {volume} {763}},\ \bibinfo {pages} {L20} (\bibinfo {year} {2013})}\BibitemShut {NoStop}%
\bibitem [{\citenamefont {{Chang}}\ \emph {et~al.}(2010)\citenamefont {{Chang}}, \citenamefont {{Pen}}, \citenamefont {{Bandura}},\ and\ \citenamefont {{Peterson}}}]{Chang_2010}%
  \BibitemOpen
  \bibfield  {author} {\bibinfo {author} {\bibfnamefont {T.-C.}\ \bibnamefont {{Chang}}}, \bibinfo {author} {\bibfnamefont {U.-L.}\ \bibnamefont {{Pen}}}, \bibinfo {author} {\bibfnamefont {K.}~\bibnamefont {{Bandura}}},\ and\ \bibinfo {author} {\bibfnamefont {J.~B.}\ \bibnamefont {{Peterson}}},\ }\bibfield  {title} {\bibinfo {title} {{An intensity map of hydrogen 21-cm emission at redshift z\raisebox{-0.5ex}\textasciitilde0.8}},\ }\href {https://doi.org/10.1038/nature09187} {\bibfield  {journal} {\bibinfo  {journal} {\nat}\ }\textbf {\bibinfo {volume} {466}},\ \bibinfo {pages} {463} (\bibinfo {year} {2010})}\BibitemShut {NoStop}%
\bibitem [{\citenamefont {Switzer}\ \emph {et~al.}(2013)\citenamefont {Switzer}, \citenamefont {Masui}, \citenamefont {Bandura}, \citenamefont {Calin}, \citenamefont {Chang}, \citenamefont {Chen}, \citenamefont {Li}, \citenamefont {Liao}, \citenamefont {Natarajan}, \citenamefont {Pen}, \citenamefont {Peterson}, \citenamefont {Shaw},\ and\ \citenamefont {Voytek}}]{Switzer_2013}%
  \BibitemOpen
  \bibfield  {author} {\bibinfo {author} {\bibfnamefont {E.~R.}\ \bibnamefont {Switzer}}, \bibinfo {author} {\bibfnamefont {K.~W.}\ \bibnamefont {Masui}}, \bibinfo {author} {\bibfnamefont {K.}~\bibnamefont {Bandura}}, \bibinfo {author} {\bibfnamefont {L.-M.}\ \bibnamefont {Calin}}, \bibinfo {author} {\bibfnamefont {T.-C.}\ \bibnamefont {Chang}}, \bibinfo {author} {\bibfnamefont {X.-L.}\ \bibnamefont {Chen}}, \bibinfo {author} {\bibfnamefont {Y.-C.}\ \bibnamefont {Li}}, \bibinfo {author} {\bibfnamefont {Y.-W.}\ \bibnamefont {Liao}}, \bibinfo {author} {\bibfnamefont {A.}~\bibnamefont {Natarajan}}, \bibinfo {author} {\bibfnamefont {U.-L.}\ \bibnamefont {Pen}}, \bibinfo {author} {\bibfnamefont {J.~B.}\ \bibnamefont {Peterson}}, \bibinfo {author} {\bibfnamefont {J.~R.}\ \bibnamefont {Shaw}},\ and\ \bibinfo {author} {\bibfnamefont {T.~C.}\ \bibnamefont {Voytek}},\ }\bibfield  {title} {\bibinfo {title} {Determination of z ~ 0.8 neutral hydrogen fluctuations using the 21 cm intensity mapping autocorrelation},\ }\href
  {https://doi.org/10.1093/mnrasl/slt074} {\bibfield  {journal} {\bibinfo  {journal} {Monthly Notices of the Royal Astronomical Society: Letters}\ }\textbf {\bibinfo {volume} {434}},\ \bibinfo {pages} {L46–L50} (\bibinfo {year} {2013})}\BibitemShut {NoStop}%
\bibitem [{\citenamefont {Abdurashidova}\ \emph {et~al.}(2022)\citenamefont {Abdurashidova}, \citenamefont {Aguirre}, \citenamefont {Alexander}, \citenamefont {Ali}, \citenamefont {Balfour}, \citenamefont {Beardsley}, \citenamefont {Bernardi}, \citenamefont {Billings}, \citenamefont {Bowman}, \citenamefont {Bradley}, \citenamefont {Bull}, \citenamefont {Burba}, \citenamefont {Carey}, \citenamefont {Carilli}, \citenamefont {Cheng}, \citenamefont {DeBoer}, \citenamefont {Dexter}, \citenamefont {de~Lera~Acedo}, \citenamefont {Dibblee-Barkman}, \citenamefont {Dillon}, \citenamefont {Ely}, \citenamefont {Ewall-Wice}, \citenamefont {Fagnoni}, \citenamefont {Fritz}, \citenamefont {Furlanetto}, \citenamefont {Gale-Sides}, \citenamefont {Glendenning}, \citenamefont {Gorthi}, \citenamefont {Greig}, \citenamefont {Grobbelaar}, \citenamefont {Halday}, \citenamefont {Hazelton}, \citenamefont {Hewitt}, \citenamefont {Hickish}, \citenamefont {Jacobs}, \citenamefont {Julius}, \citenamefont {Kern}, \citenamefont {Kerrigan},
  \citenamefont {Kittiwisit}, \citenamefont {Kohn}, \citenamefont {Kolopanis}, \citenamefont {Lanman}, \citenamefont {La~Plante}, \citenamefont {Lekalake}, \citenamefont {Lewis}, \citenamefont {Liu}, \citenamefont {MacMahon}, \citenamefont {Malan}, \citenamefont {Malgas}, \citenamefont {Maree}, \citenamefont {Martinot}, \citenamefont {Matsetela}, \citenamefont {Mesinger}, \citenamefont {Molewa}, \citenamefont {Morales}, \citenamefont {Mosiane}, \citenamefont {Murray}, \citenamefont {Neben}, \citenamefont {Nikolic}, \citenamefont {Nunhokee}, \citenamefont {Parsons}, \citenamefont {Patra}, \citenamefont {Pascua}, \citenamefont {Pieterse}, \citenamefont {Pober}, \citenamefont {Razavi-Ghods}, \citenamefont {Ringuette}, \citenamefont {Robnett}, \citenamefont {Rosie}, \citenamefont {Sims}, \citenamefont {Singh}, \citenamefont {Smith}, \citenamefont {Syce}, \citenamefont {Thyagarajan}, \citenamefont {Williams},\ and\ \citenamefont {Zheng}}]{HERA_limit}%
  \BibitemOpen
  \bibfield  {author} {\bibinfo {author} {\bibfnamefont {Z.}~\bibnamefont {Abdurashidova}}, \bibinfo {author} {\bibfnamefont {J.~E.}\ \bibnamefont {Aguirre}}, \bibinfo {author} {\bibfnamefont {P.}~\bibnamefont {Alexander}}, \bibinfo {author} {\bibfnamefont {Z.~S.}\ \bibnamefont {Ali}}, \bibinfo {author} {\bibfnamefont {Y.}~\bibnamefont {Balfour}}, \bibinfo {author} {\bibfnamefont {A.~P.}\ \bibnamefont {Beardsley}}, \bibinfo {author} {\bibfnamefont {G.}~\bibnamefont {Bernardi}}, \bibinfo {author} {\bibfnamefont {T.~S.}\ \bibnamefont {Billings}}, \bibinfo {author} {\bibfnamefont {J.~D.}\ \bibnamefont {Bowman}}, \bibinfo {author} {\bibfnamefont {R.~F.}\ \bibnamefont {Bradley}}, \bibinfo {author} {\bibfnamefont {P.}~\bibnamefont {Bull}}, \bibinfo {author} {\bibfnamefont {J.}~\bibnamefont {Burba}}, \bibinfo {author} {\bibfnamefont {S.}~\bibnamefont {Carey}}, \bibinfo {author} {\bibfnamefont {C.~L.}\ \bibnamefont {Carilli}}, \bibinfo {author} {\bibfnamefont {C.}~\bibnamefont {Cheng}}, \bibinfo {author}
  {\bibfnamefont {D.~R.}\ \bibnamefont {DeBoer}}, \bibinfo {author} {\bibfnamefont {M.}~\bibnamefont {Dexter}}, \bibinfo {author} {\bibfnamefont {E.}~\bibnamefont {de~Lera~Acedo}}, \bibinfo {author} {\bibfnamefont {T.}~\bibnamefont {Dibblee-Barkman}}, \bibinfo {author} {\bibfnamefont {J.~S.}\ \bibnamefont {Dillon}}, \bibinfo {author} {\bibfnamefont {J.}~\bibnamefont {Ely}}, \bibinfo {author} {\bibfnamefont {A.}~\bibnamefont {Ewall-Wice}}, \bibinfo {author} {\bibfnamefont {N.}~\bibnamefont {Fagnoni}}, \bibinfo {author} {\bibfnamefont {R.}~\bibnamefont {Fritz}}, \bibinfo {author} {\bibfnamefont {S.~R.}\ \bibnamefont {Furlanetto}}, \bibinfo {author} {\bibfnamefont {K.}~\bibnamefont {Gale-Sides}}, \bibinfo {author} {\bibfnamefont {B.}~\bibnamefont {Glendenning}}, \bibinfo {author} {\bibfnamefont {D.}~\bibnamefont {Gorthi}}, \bibinfo {author} {\bibfnamefont {B.}~\bibnamefont {Greig}}, \bibinfo {author} {\bibfnamefont {J.}~\bibnamefont {Grobbelaar}}, \bibinfo {author} {\bibfnamefont {Z.}~\bibnamefont {Halday}},
  \bibinfo {author} {\bibfnamefont {B.~J.}\ \bibnamefont {Hazelton}}, \bibinfo {author} {\bibfnamefont {J.~N.}\ \bibnamefont {Hewitt}}, \bibinfo {author} {\bibfnamefont {J.}~\bibnamefont {Hickish}}, \bibinfo {author} {\bibfnamefont {D.~C.}\ \bibnamefont {Jacobs}}, \bibinfo {author} {\bibfnamefont {A.}~\bibnamefont {Julius}}, \bibinfo {author} {\bibfnamefont {N.~S.}\ \bibnamefont {Kern}}, \bibinfo {author} {\bibfnamefont {J.}~\bibnamefont {Kerrigan}}, \bibinfo {author} {\bibfnamefont {P.}~\bibnamefont {Kittiwisit}}, \bibinfo {author} {\bibfnamefont {S.~A.}\ \bibnamefont {Kohn}}, \bibinfo {author} {\bibfnamefont {M.}~\bibnamefont {Kolopanis}}, \bibinfo {author} {\bibfnamefont {A.}~\bibnamefont {Lanman}}, \bibinfo {author} {\bibfnamefont {P.}~\bibnamefont {La~Plante}}, \bibinfo {author} {\bibfnamefont {T.}~\bibnamefont {Lekalake}}, \bibinfo {author} {\bibfnamefont {D.}~\bibnamefont {Lewis}}, \bibinfo {author} {\bibfnamefont {A.}~\bibnamefont {Liu}}, \bibinfo {author} {\bibfnamefont {D.}~\bibnamefont {MacMahon}},
  \bibinfo {author} {\bibfnamefont {L.}~\bibnamefont {Malan}}, \bibinfo {author} {\bibfnamefont {C.}~\bibnamefont {Malgas}}, \bibinfo {author} {\bibfnamefont {M.}~\bibnamefont {Maree}}, \bibinfo {author} {\bibfnamefont {Z.~E.}\ \bibnamefont {Martinot}}, \bibinfo {author} {\bibfnamefont {E.}~\bibnamefont {Matsetela}}, \bibinfo {author} {\bibfnamefont {A.}~\bibnamefont {Mesinger}}, \bibinfo {author} {\bibfnamefont {M.}~\bibnamefont {Molewa}}, \bibinfo {author} {\bibfnamefont {M.~F.}\ \bibnamefont {Morales}}, \bibinfo {author} {\bibfnamefont {T.}~\bibnamefont {Mosiane}}, \bibinfo {author} {\bibfnamefont {S.~G.}\ \bibnamefont {Murray}}, \bibinfo {author} {\bibfnamefont {A.~R.}\ \bibnamefont {Neben}}, \bibinfo {author} {\bibfnamefont {B.}~\bibnamefont {Nikolic}}, \bibinfo {author} {\bibfnamefont {C.~D.}\ \bibnamefont {Nunhokee}}, \bibinfo {author} {\bibfnamefont {A.~R.}\ \bibnamefont {Parsons}}, \bibinfo {author} {\bibfnamefont {N.}~\bibnamefont {Patra}}, \bibinfo {author} {\bibfnamefont {R.}~\bibnamefont
  {Pascua}}, \bibinfo {author} {\bibfnamefont {S.}~\bibnamefont {Pieterse}}, \bibinfo {author} {\bibfnamefont {J.~C.}\ \bibnamefont {Pober}}, \bibinfo {author} {\bibfnamefont {N.}~\bibnamefont {Razavi-Ghods}}, \bibinfo {author} {\bibfnamefont {J.}~\bibnamefont {Ringuette}}, \bibinfo {author} {\bibfnamefont {J.}~\bibnamefont {Robnett}}, \bibinfo {author} {\bibfnamefont {K.}~\bibnamefont {Rosie}}, \bibinfo {author} {\bibfnamefont {P.}~\bibnamefont {Sims}}, \bibinfo {author} {\bibfnamefont {S.}~\bibnamefont {Singh}}, \bibinfo {author} {\bibfnamefont {C.}~\bibnamefont {Smith}}, \bibinfo {author} {\bibfnamefont {A.}~\bibnamefont {Syce}}, \bibinfo {author} {\bibfnamefont {N.}~\bibnamefont {Thyagarajan}}, \bibinfo {author} {\bibfnamefont {P.~K.~G.}\ \bibnamefont {Williams}},\ and\ \bibinfo {author} {\bibfnamefont {H.}~\bibnamefont {Zheng}},\ }\bibfield  {title} {\bibinfo {title} {First results from hera phase i: Upper limits on the epoch of reionization 21 cm power spectrum},\ }\href
  {https://doi.org/10.3847/1538-4357/ac1c78} {\bibfield  {journal} {\bibinfo  {journal} {The Astrophysical Journal}\ }\textbf {\bibinfo {volume} {925}},\ \bibinfo {pages} {221} (\bibinfo {year} {2022})}\BibitemShut {NoStop}%
\bibitem [{\citenamefont {Paul}\ \emph {et~al.}(2023)\citenamefont {Paul}, \citenamefont {Santos}, \citenamefont {Chen},\ and\ \citenamefont {Wolz}}]{paul2023detection}%
  \BibitemOpen
  \bibfield  {author} {\bibinfo {author} {\bibfnamefont {S.}~\bibnamefont {Paul}}, \bibinfo {author} {\bibfnamefont {M.~G.}\ \bibnamefont {Santos}}, \bibinfo {author} {\bibfnamefont {Z.}~\bibnamefont {Chen}},\ and\ \bibinfo {author} {\bibfnamefont {L.}~\bibnamefont {Wolz}},\ }\href@noop {} {\bibinfo {title} {A first detection of neutral hydrogen intensity mapping on mpc scales at $z\approx 0.32$ and $z\approx 0.44$}} (\bibinfo {year} {2023}),\ \Eprint {https://arxiv.org/abs/2301.11943} {arXiv:2301.11943 [astro-ph.CO]} \BibitemShut {NoStop}%
\bibitem [{\citenamefont {Shaw}\ \emph {et~al.}(2014)\citenamefont {Shaw}, \citenamefont {Sigurdson}, \citenamefont {Pen}, \citenamefont {Stebbins},\ and\ \citenamefont {Sitwell}}]{first_mmode}%
  \BibitemOpen
  \bibfield  {author} {\bibinfo {author} {\bibfnamefont {J.~R.}\ \bibnamefont {Shaw}}, \bibinfo {author} {\bibfnamefont {K.}~\bibnamefont {Sigurdson}}, \bibinfo {author} {\bibfnamefont {U.-L.}\ \bibnamefont {Pen}}, \bibinfo {author} {\bibfnamefont {A.}~\bibnamefont {Stebbins}},\ and\ \bibinfo {author} {\bibfnamefont {M.}~\bibnamefont {Sitwell}},\ }\bibfield  {title} {\bibinfo {title} {All-sky interferometry with spherical harmonic transit telescopes},\ }\href {https://doi.org/10.1088/0004-637x/781/2/57} {\bibfield  {journal} {\bibinfo  {journal} {The Astrophysical Journal}\ }\textbf {\bibinfo {volume} {781}},\ \bibinfo {pages} {57} (\bibinfo {year} {2014})}\BibitemShut {NoStop}%
\bibitem [{\citenamefont {Ewall-Wice}\ \emph {et~al.}(2020)\citenamefont {Ewall-Wice}, \citenamefont {Kern}, \citenamefont {Dillon}, \citenamefont {Liu}, \citenamefont {Parsons}, \citenamefont {Singh}, \citenamefont {Lanman}, \citenamefont {Plante}, \citenamefont {Fagnoni}, \citenamefont {Acedo}, \citenamefont {DeBoer}, \citenamefont {Nunhokee}, \citenamefont {Bull}, \citenamefont {Chang}, \citenamefont {Lazio}, \citenamefont {Aguirre},\ and\ \citenamefont {Weinberg}}]{DAYENU}%
  \BibitemOpen
  \bibfield  {author} {\bibinfo {author} {\bibfnamefont {A.}~\bibnamefont {Ewall-Wice}}, \bibinfo {author} {\bibfnamefont {N.}~\bibnamefont {Kern}}, \bibinfo {author} {\bibfnamefont {J.~S.}\ \bibnamefont {Dillon}}, \bibinfo {author} {\bibfnamefont {A.}~\bibnamefont {Liu}}, \bibinfo {author} {\bibfnamefont {A.}~\bibnamefont {Parsons}}, \bibinfo {author} {\bibfnamefont {S.}~\bibnamefont {Singh}}, \bibinfo {author} {\bibfnamefont {A.}~\bibnamefont {Lanman}}, \bibinfo {author} {\bibfnamefont {P.~L.}\ \bibnamefont {Plante}}, \bibinfo {author} {\bibfnamefont {N.}~\bibnamefont {Fagnoni}}, \bibinfo {author} {\bibfnamefont {E.~d.~L.}\ \bibnamefont {Acedo}}, \bibinfo {author} {\bibfnamefont {D.~R.}\ \bibnamefont {DeBoer}}, \bibinfo {author} {\bibfnamefont {C.}~\bibnamefont {Nunhokee}}, \bibinfo {author} {\bibfnamefont {P.}~\bibnamefont {Bull}}, \bibinfo {author} {\bibfnamefont {T.-C.}\ \bibnamefont {Chang}}, \bibinfo {author} {\bibfnamefont {T.~J.~W.}\ \bibnamefont {Lazio}}, \bibinfo {author} {\bibfnamefont
  {J.}~\bibnamefont {Aguirre}},\ and\ \bibinfo {author} {\bibfnamefont {S.}~\bibnamefont {Weinberg}},\ }\bibfield  {title} {\bibinfo {title} {{DAYENU: a simple filter of smooth foregrounds for intensity mapping power spectra}},\ }\href {https://doi.org/10.1093/mnras/staa3293} {\bibfield  {journal} {\bibinfo  {journal} {Monthly Notices of the Royal Astronomical Society}\ }\textbf {\bibinfo {volume} {500}},\ \bibinfo {pages} {5195} (\bibinfo {year} {2020})},\ \Eprint {https://arxiv.org/abs/https://academic.oup.com/mnras/article-pdf/500/4/5195/34908165/staa3293.pdf} {https://academic.oup.com/mnras/article-pdf/500/4/5195/34908165/staa3293.pdf} \BibitemShut {NoStop}%
\bibitem [{\citenamefont {Bond}(1995)}]{KL_cite}%
  \BibitemOpen
  \bibfield  {author} {\bibinfo {author} {\bibfnamefont {J.~R.}\ \bibnamefont {Bond}},\ }\bibfield  {title} {\bibinfo {title} {Signal-to-noise eigenmode analysis of the two-year cobe maps},\ }\href {https://doi.org/10.1103/PhysRevLett.74.4369} {\bibfield  {journal} {\bibinfo  {journal} {Phys. Rev. Lett.}\ }\textbf {\bibinfo {volume} {74}},\ \bibinfo {pages} {4369} (\bibinfo {year} {1995})}\BibitemShut {NoStop}%
\bibitem [{\citenamefont {Shaw}\ \emph {et~al.}(2015)\citenamefont {Shaw}, \citenamefont {Sigurdson}, \citenamefont {Sitwell}, \citenamefont {Stebbins},\ and\ \citenamefont {Pen}}]{mmode}%
  \BibitemOpen
  \bibfield  {author} {\bibinfo {author} {\bibfnamefont {J.~R.}\ \bibnamefont {Shaw}}, \bibinfo {author} {\bibfnamefont {K.}~\bibnamefont {Sigurdson}}, \bibinfo {author} {\bibfnamefont {M.}~\bibnamefont {Sitwell}}, \bibinfo {author} {\bibfnamefont {A.}~\bibnamefont {Stebbins}},\ and\ \bibinfo {author} {\bibfnamefont {U.-L.}\ \bibnamefont {Pen}},\ }\bibfield  {title} {\bibinfo {title} {Coaxing cosmic 21 cm fluctuations from the polarized sky using $m$-mode analysis},\ }\href {https://doi.org/10.1103/PhysRevD.91.083514} {\bibfield  {journal} {\bibinfo  {journal} {Phys. Rev. D}\ }\textbf {\bibinfo {volume} {91}},\ \bibinfo {pages} {083514} (\bibinfo {year} {2015})}\BibitemShut {NoStop}%
\bibitem [{\citenamefont {Wang}\ \emph {et~al.}(2022)\citenamefont {Wang}, \citenamefont {Mena-Parra}, \citenamefont {Chen},\ and\ \citenamefont {Masui}}]{haochen_first_paper}%
  \BibitemOpen
  \bibfield  {author} {\bibinfo {author} {\bibfnamefont {H.}~\bibnamefont {Wang}}, \bibinfo {author} {\bibfnamefont {J.}~\bibnamefont {Mena-Parra}}, \bibinfo {author} {\bibfnamefont {T.}~\bibnamefont {Chen}},\ and\ \bibinfo {author} {\bibfnamefont {K.}~\bibnamefont {Masui}},\ }\bibfield  {title} {\bibinfo {title} {Removing systematics-induced 21-cm foreground residuals by cross-correlating filtered data},\ }\href {https://doi.org/10.1103/PhysRevD.106.043534} {\bibfield  {journal} {\bibinfo  {journal} {Phys. Rev. D}\ }\textbf {\bibinfo {volume} {106}},\ \bibinfo {pages} {043534} (\bibinfo {year} {2022})}\BibitemShut {NoStop}%
\bibitem [{\citenamefont {Ross}\ \emph {et~al.}(2020)\citenamefont {Ross}, \citenamefont {Bautista}, \citenamefont {Tojeiro}, \citenamefont {Alam}, \citenamefont {Bailey}, \citenamefont {Burtin}, \citenamefont {Comparat}, \citenamefont {Dawson}, \citenamefont {de~Mattia}, \citenamefont {du~Mas~des Bourboux}, \citenamefont {Gil-Marín}, \citenamefont {Hou}, \citenamefont {Kong}, \citenamefont {Lyke}, \citenamefont {Mohammad}, \citenamefont {Moustakas}, \citenamefont {Mueller}, \citenamefont {Myers}, \citenamefont {Percival}, \citenamefont {Raichoor}, \citenamefont {Rezaie}, \citenamefont {Seo}, \citenamefont {Smith}, \citenamefont {Tinker}, \citenamefont {Zarrouk}, \citenamefont {Zhao}, \citenamefont {Zhao}, \citenamefont {Bizyaev}, \citenamefont {Brinkmann}, \citenamefont {Brownstein}, \citenamefont {Rosell}, \citenamefont {Chabanier}, \citenamefont {Choi}, \citenamefont {Chuang}, \citenamefont {Cruz-Gonzalez}, \citenamefont {de~la Macorra}, \citenamefont {de~la Torre}, \citenamefont {Escoffier},
  \citenamefont {Fromenteau}, \citenamefont {Higley}, \citenamefont {Jullo}, \citenamefont {Kneib}, \citenamefont {McLane}, \citenamefont {Muñoz-Gutiérrez}, \citenamefont {Neveux}, \citenamefont {Newman}, \citenamefont {Nitschelm}, \citenamefont {Palanque-Delabrouille}, \citenamefont {Paviot}, \citenamefont {Pullen}, \citenamefont {Rossi}, \citenamefont {Ruhlmann-Kleider}, \citenamefont {Schneider}, \citenamefont {Magaña}, \citenamefont {Vivek},\ and\ \citenamefont {Zhang}}]{eboss_rand_1}%
  \BibitemOpen
  \bibfield  {author} {\bibinfo {author} {\bibfnamefont {A.~J.}\ \bibnamefont {Ross}}, \bibinfo {author} {\bibfnamefont {J.}~\bibnamefont {Bautista}}, \bibinfo {author} {\bibfnamefont {R.}~\bibnamefont {Tojeiro}}, \bibinfo {author} {\bibfnamefont {S.}~\bibnamefont {Alam}}, \bibinfo {author} {\bibfnamefont {S.}~\bibnamefont {Bailey}}, \bibinfo {author} {\bibfnamefont {E.}~\bibnamefont {Burtin}}, \bibinfo {author} {\bibfnamefont {J.}~\bibnamefont {Comparat}}, \bibinfo {author} {\bibfnamefont {K.~S.}\ \bibnamefont {Dawson}}, \bibinfo {author} {\bibfnamefont {A.}~\bibnamefont {de~Mattia}}, \bibinfo {author} {\bibfnamefont {H.}~\bibnamefont {du~Mas~des Bourboux}}, \bibinfo {author} {\bibfnamefont {H.}~\bibnamefont {Gil-Marín}}, \bibinfo {author} {\bibfnamefont {J.}~\bibnamefont {Hou}}, \bibinfo {author} {\bibfnamefont {H.}~\bibnamefont {Kong}}, \bibinfo {author} {\bibfnamefont {B.~W.}\ \bibnamefont {Lyke}}, \bibinfo {author} {\bibfnamefont {F.~G.}\ \bibnamefont {Mohammad}}, \bibinfo {author} {\bibfnamefont
  {J.}~\bibnamefont {Moustakas}}, \bibinfo {author} {\bibfnamefont {E.-M.}\ \bibnamefont {Mueller}}, \bibinfo {author} {\bibfnamefont {A.~D.}\ \bibnamefont {Myers}}, \bibinfo {author} {\bibfnamefont {W.~J.}\ \bibnamefont {Percival}}, \bibinfo {author} {\bibfnamefont {A.}~\bibnamefont {Raichoor}}, \bibinfo {author} {\bibfnamefont {M.}~\bibnamefont {Rezaie}}, \bibinfo {author} {\bibfnamefont {H.-J.}\ \bibnamefont {Seo}}, \bibinfo {author} {\bibfnamefont {A.}~\bibnamefont {Smith}}, \bibinfo {author} {\bibfnamefont {J.~L.}\ \bibnamefont {Tinker}}, \bibinfo {author} {\bibfnamefont {P.}~\bibnamefont {Zarrouk}}, \bibinfo {author} {\bibfnamefont {C.}~\bibnamefont {Zhao}}, \bibinfo {author} {\bibfnamefont {G.-B.}\ \bibnamefont {Zhao}}, \bibinfo {author} {\bibfnamefont {D.}~\bibnamefont {Bizyaev}}, \bibinfo {author} {\bibfnamefont {J.}~\bibnamefont {Brinkmann}}, \bibinfo {author} {\bibfnamefont {J.~R.}\ \bibnamefont {Brownstein}}, \bibinfo {author} {\bibfnamefont {A.~C.}\ \bibnamefont {Rosell}}, \bibinfo {author}
  {\bibfnamefont {S.}~\bibnamefont {Chabanier}}, \bibinfo {author} {\bibfnamefont {P.~D.}\ \bibnamefont {Choi}}, \bibinfo {author} {\bibfnamefont {C.-H.}\ \bibnamefont {Chuang}}, \bibinfo {author} {\bibfnamefont {I.}~\bibnamefont {Cruz-Gonzalez}}, \bibinfo {author} {\bibfnamefont {A.}~\bibnamefont {de~la Macorra}}, \bibinfo {author} {\bibfnamefont {S.}~\bibnamefont {de~la Torre}}, \bibinfo {author} {\bibfnamefont {S.}~\bibnamefont {Escoffier}}, \bibinfo {author} {\bibfnamefont {S.}~\bibnamefont {Fromenteau}}, \bibinfo {author} {\bibfnamefont {A.}~\bibnamefont {Higley}}, \bibinfo {author} {\bibfnamefont {E.}~\bibnamefont {Jullo}}, \bibinfo {author} {\bibfnamefont {J.-P.}\ \bibnamefont {Kneib}}, \bibinfo {author} {\bibfnamefont {J.~N.}\ \bibnamefont {McLane}}, \bibinfo {author} {\bibfnamefont {A.}~\bibnamefont {Muñoz-Gutiérrez}}, \bibinfo {author} {\bibfnamefont {R.}~\bibnamefont {Neveux}}, \bibinfo {author} {\bibfnamefont {J.~A.}\ \bibnamefont {Newman}}, \bibinfo {author} {\bibfnamefont {C.}~\bibnamefont
  {Nitschelm}}, \bibinfo {author} {\bibfnamefont {N.}~\bibnamefont {Palanque-Delabrouille}}, \bibinfo {author} {\bibfnamefont {R.}~\bibnamefont {Paviot}}, \bibinfo {author} {\bibfnamefont {A.~R.}\ \bibnamefont {Pullen}}, \bibinfo {author} {\bibfnamefont {G.}~\bibnamefont {Rossi}}, \bibinfo {author} {\bibfnamefont {V.}~\bibnamefont {Ruhlmann-Kleider}}, \bibinfo {author} {\bibfnamefont {D.~P.}\ \bibnamefont {Schneider}}, \bibinfo {author} {\bibfnamefont {M.~V.}\ \bibnamefont {Magaña}}, \bibinfo {author} {\bibfnamefont {M.}~\bibnamefont {Vivek}},\ and\ \bibinfo {author} {\bibfnamefont {Y.}~\bibnamefont {Zhang}},\ }\bibfield  {title} {\bibinfo {title} {The completed sdss-iv extended baryon oscillation spectroscopic survey: Large-scale structure catalogues for cosmological analysis},\ }\href {https://doi.org/10.1093/mnras/staa2416} {\bibfield  {journal} {\bibinfo  {journal} {Monthly Notices of the Royal Astronomical Society}\ }\textbf {\bibinfo {volume} {498}},\ \bibinfo {pages} {2354–2371} (\bibinfo {year}
  {2020})}\BibitemShut {NoStop}%
\bibitem [{\citenamefont {Raichoor}\ \emph {et~al.}(2020)\citenamefont {Raichoor}, \citenamefont {de~Mattia}, \citenamefont {Ross}, \citenamefont {Zhao}, \citenamefont {Alam}, \citenamefont {Avila}, \citenamefont {Bautista}, \citenamefont {Brinkmann}, \citenamefont {Brownstein}, \citenamefont {Burtin}, \citenamefont {Chapman}, \citenamefont {Chuang}, \citenamefont {Comparat}, \citenamefont {Dawson}, \citenamefont {Dey}, \citenamefont {du~Mas~des Bourboux}, \citenamefont {Elvin-Poole}, \citenamefont {Gonzalez-Perez}, \citenamefont {Gorgoni}, \citenamefont {Kneib}, \citenamefont {Kong}, \citenamefont {Lang}, \citenamefont {Moustakas}, \citenamefont {Myers}, \citenamefont {Müller}, \citenamefont {Nadathur}, \citenamefont {Newman}, \citenamefont {Percival}, \citenamefont {Rezaie}, \citenamefont {Rossi}, \citenamefont {Ruhlmann-Kleider}, \citenamefont {Schlegel}, \citenamefont {Schneider}, \citenamefont {Seo}, \citenamefont {Tamone}, \citenamefont {Tinker}, \citenamefont {Tojeiro}, \citenamefont {Vivek},
  \citenamefont {Yèche},\ and\ \citenamefont {Zhao}}]{eboss_rand_2}%
  \BibitemOpen
  \bibfield  {author} {\bibinfo {author} {\bibfnamefont {A.}~\bibnamefont {Raichoor}}, \bibinfo {author} {\bibfnamefont {A.}~\bibnamefont {de~Mattia}}, \bibinfo {author} {\bibfnamefont {A.~J.}\ \bibnamefont {Ross}}, \bibinfo {author} {\bibfnamefont {C.}~\bibnamefont {Zhao}}, \bibinfo {author} {\bibfnamefont {S.}~\bibnamefont {Alam}}, \bibinfo {author} {\bibfnamefont {S.}~\bibnamefont {Avila}}, \bibinfo {author} {\bibfnamefont {J.}~\bibnamefont {Bautista}}, \bibinfo {author} {\bibfnamefont {J.}~\bibnamefont {Brinkmann}}, \bibinfo {author} {\bibfnamefont {J.~R.}\ \bibnamefont {Brownstein}}, \bibinfo {author} {\bibfnamefont {E.}~\bibnamefont {Burtin}}, \bibinfo {author} {\bibfnamefont {M.~J.}\ \bibnamefont {Chapman}}, \bibinfo {author} {\bibfnamefont {C.-H.}\ \bibnamefont {Chuang}}, \bibinfo {author} {\bibfnamefont {J.}~\bibnamefont {Comparat}}, \bibinfo {author} {\bibfnamefont {K.~S.}\ \bibnamefont {Dawson}}, \bibinfo {author} {\bibfnamefont {A.}~\bibnamefont {Dey}}, \bibinfo {author} {\bibfnamefont
  {H.}~\bibnamefont {du~Mas~des Bourboux}}, \bibinfo {author} {\bibfnamefont {J.}~\bibnamefont {Elvin-Poole}}, \bibinfo {author} {\bibfnamefont {V.}~\bibnamefont {Gonzalez-Perez}}, \bibinfo {author} {\bibfnamefont {C.}~\bibnamefont {Gorgoni}}, \bibinfo {author} {\bibfnamefont {J.-P.}\ \bibnamefont {Kneib}}, \bibinfo {author} {\bibfnamefont {H.}~\bibnamefont {Kong}}, \bibinfo {author} {\bibfnamefont {D.}~\bibnamefont {Lang}}, \bibinfo {author} {\bibfnamefont {J.}~\bibnamefont {Moustakas}}, \bibinfo {author} {\bibfnamefont {A.~D.}\ \bibnamefont {Myers}}, \bibinfo {author} {\bibfnamefont {E.-M.}\ \bibnamefont {Müller}}, \bibinfo {author} {\bibfnamefont {S.}~\bibnamefont {Nadathur}}, \bibinfo {author} {\bibfnamefont {J.~A.}\ \bibnamefont {Newman}}, \bibinfo {author} {\bibfnamefont {W.~J.}\ \bibnamefont {Percival}}, \bibinfo {author} {\bibfnamefont {M.}~\bibnamefont {Rezaie}}, \bibinfo {author} {\bibfnamefont {G.}~\bibnamefont {Rossi}}, \bibinfo {author} {\bibfnamefont {V.}~\bibnamefont {Ruhlmann-Kleider}},
  \bibinfo {author} {\bibfnamefont {D.~J.}\ \bibnamefont {Schlegel}}, \bibinfo {author} {\bibfnamefont {D.~P.}\ \bibnamefont {Schneider}}, \bibinfo {author} {\bibfnamefont {H.-J.}\ \bibnamefont {Seo}}, \bibinfo {author} {\bibfnamefont {A.}~\bibnamefont {Tamone}}, \bibinfo {author} {\bibfnamefont {J.~L.}\ \bibnamefont {Tinker}}, \bibinfo {author} {\bibfnamefont {R.}~\bibnamefont {Tojeiro}}, \bibinfo {author} {\bibfnamefont {M.}~\bibnamefont {Vivek}}, \bibinfo {author} {\bibfnamefont {C.}~\bibnamefont {Yèche}},\ and\ \bibinfo {author} {\bibfnamefont {G.-B.}\ \bibnamefont {Zhao}},\ }\bibfield  {title} {\bibinfo {title} {The completed sdss-iv extended baryon oscillation spectroscopic survey: large-scale structure catalogues and measurement of the isotropic bao between redshift 0.6 and 1.1 for the emission line galaxy sample},\ }\href {https://doi.org/10.1093/mnras/staa3336} {\bibfield  {journal} {\bibinfo  {journal} {Monthly Notices of the Royal Astronomical Society}\ }\textbf {\bibinfo {volume} {500}},\
  \bibinfo {pages} {3254–3274} (\bibinfo {year} {2020})}\BibitemShut {NoStop}%
\bibitem [{\citenamefont {Mead}\ \emph {et~al.}(2021)\citenamefont {Mead}, \citenamefont {Brieden}, \citenamefont {Tröster},\ and\ \citenamefont {Heymans}}]{Mead_2021}%
  \BibitemOpen
  \bibfield  {author} {\bibinfo {author} {\bibfnamefont {A.~J.}\ \bibnamefont {Mead}}, \bibinfo {author} {\bibfnamefont {S.}~\bibnamefont {Brieden}}, \bibinfo {author} {\bibfnamefont {T.}~\bibnamefont {Tröster}},\ and\ \bibinfo {author} {\bibfnamefont {C.}~\bibnamefont {Heymans}},\ }\bibfield  {title} {\bibinfo {title} {<scp>hmcode-2020</scp>: improved modelling of non-linear cosmological power spectra with baryonic feedback},\ }\href {https://doi.org/10.1093/mnras/stab082} {\bibfield  {journal} {\bibinfo  {journal} {Monthly Notices of the Royal Astronomical Society}\ }\textbf {\bibinfo {volume} {502}},\ \bibinfo {pages} {1401–1422} (\bibinfo {year} {2021})}\BibitemShut {NoStop}%
\bibitem [{\citenamefont {{Rybicki}}\ and\ \citenamefont {{Press}}(1992)}]{wiener_filter}%
  \BibitemOpen
  \bibfield  {author} {\bibinfo {author} {\bibfnamefont {G.~B.}\ \bibnamefont {{Rybicki}}}\ and\ \bibinfo {author} {\bibfnamefont {W.~H.}\ \bibnamefont {{Press}}},\ }\bibfield  {title} {\bibinfo {title} {{Interpolation, Realization, and Reconstruction of Noisy, Irregularly Sampled Data}},\ }\href {https://doi.org/10.1086/171845} {\bibfield  {journal} {\bibinfo  {journal} {\apj}\ }\textbf {\bibinfo {volume} {398}},\ \bibinfo {pages} {169} (\bibinfo {year} {1992})}\BibitemShut {NoStop}%
\end{thebibliography}%

\end{document}